\documentclass[twocolumn]{aastex63}
\usepackage{graphicx}
\usepackage{amsmath}
\usepackage{txfonts}

\newcommand{\nnotreject}{1095}
\newcommand{\nreliable}{349}
\newcommand{\nfull}{748} 
\newcommand{\nremovedduringfitting}{347}  

\newcommand{\ncombined}{809} 
\newcommand{\ncombinedTESS}{349} 
\newcommand{\ncombinedKEPLER}{365} 
\newcommand{\ncombinedTORRES}{95} 

\received{\today}
\revised{\today}
\accepted{\today}
\submitjournal{ApJ}

\shorttitle{Tidal Circularization of Close Binaries}
\shortauthors{Justesen \& Albrecht}

\begin{document} 
\title{Temperature and Distance Dependence of Tidal Circularization in Close Binaries: \\
A Catalog of Eclipsing Binaries in the Southern Hemisphere Observed by the TESS Satellite.}

\correspondingauthor{Anders B. Justesen}
\email{justesen@phys.au.dk}

\author[0000-0002-0174-2466]{Anders B. Justesen}
\affiliation{Stellar Astrophysics Centre, Department of Physics and Astronomy, Aarhus University, Ny Munkegade 120, 8000 Aarhus C, Denmark}

\author[0000-0003-1762-8235]{Simon Albrecht}
\affiliation{Stellar Astrophysics Centre, Department of Physics and Astronomy, Aarhus University, Ny Munkegade 120, 8000 Aarhus C, Denmark}

\begin{abstract}
Tidal forces are important for understanding how close binary stars and compact exoplanetary systems form and evolve. However, tides are difficult to model and significant uncertainties exist about the strength of tides. Here, we investigate tidal circularization in close binaries using a large sample of well-characterised eclipsing systems. We searched TESS photometry from the southern hemisphere for eclipsing binaries. We derive best-fit orbital and stellar parameters by jointly modelling light curves and spectral energy distributions. To determine the eccentricity distribution of eclipsing binaries over a wide range of stellar temperatures ($3\,000-50\,000\,$K) and orbital separations $a/R_1$ ($2-300$), we combine our newly obtained TESS sample with eclipsing binaries observed from the ground and by the \textit{Kepler} mission. We find a clear dependency of stellar temperature and orbital separation in the eccentricities of close binaries. We compare our observations with predictions of the equilibrium and dynamical tides. We find that while cool binaries agree with the predictions of the equilibrium tide, a large fraction of binaries with temperatures between $6\,250$~K and $10\,000\,$~K and orbital separations between $a/R_1\sim 4$ and $10$ are found on circular orbits contrary to the predictions of the dynamical tide. This suggests that some binaries with radiative envelopes may be tidally circularised significantly more efficiently than usually assumed. Our findings on orbital circularization have important implications also in the context of hot Jupiters where tides have been invoked to explain the observed difference in the spin-orbit alignment between hot and cool host stars.
\end{abstract}

   \keywords{Binary stars (154), Close binary stars (254), Detached binary stars (375), Eclipsing binary stars (444), Tides (1702), Tidal interaction (1699), Light curves (918), Transit photometry (1709), Orbit determination (1175), Orbital evolution (1178)}

\section{Introduction}
Binary stars, and in particular eclipsing binaries, are important for our understanding of a wealth of astrophysical topics. The study of eclipsing binaries allows the determination of absolute stellar and orbital parameters by combining photometry and radial velocities. The determination of near model-independent precise parameters are important for testing and calibrating both stellar evolutionary models \citep[e.g.][]{Higl2017, Hidalgo2018} and observational techniques such as asteroseismic scaling relations \citep[e.g.][]{Brogaard2016} and distance indicators \citep[e.g.][]{Pietrzynski2013}. Eclipsing binaries further enable the study of topics such as tidal interactions \citep[e.g. tidally trapped pulsations,][]{Handler2020}, tidal circularization \citep[e.g.][]{Meibom2005}, tidal spin-orbit alignment \cite[e.g.][]{Albrecht2009}, and orbital decay \citep[e.g. testing general relativistic predictions,][]{Burdge2019}. Studies of large samples of binary stars can improve our understanding stellar formation \citep[e.g.][]{White2001}, stellar populations \citep[e.g.][]{Stanway2018}, exoplanet demographics \citep[e.g.][]{Ciardi2015} and galactic and extragalatic astronomy \citep[e.g.][]{Kroupa2001, Riess2016}.

The advent of large surveys with continuous monitoring of the sky has enabled the discovery and characterisation of a large number of eclipsing binaries. Thousands of eclipsing binaries have been detected in the Magellanic Clouds by the Optical Gravitational Lensing Experiment \citep[OGLE;][]{Udalski1998, Wyrzykowski2004} and MACHO project \citep{Alcock1997, Derekas2007}. This data have been used to investigate tidal circularization among B-type stars \citep[]{North2003, Mazeh2006}. The survey of eclipsing binaries from the Trans-Atlantic Exoplanet Survey \citep[TrES;][]{Alonso2004} was similarly used to study the period-eccentricity distribution of eclipsing binaries \citep{Devor2008}. \citet{Torres2010} compiled a list of well-detached eclipsing binaries with accurate fundamental parameters and reported a difference in the eccentricity distribution of early- and late-type binaries, as expected from tidal theory. The \textit{Kepler} mission \citep{Borucki2010} detected thousands of eclipsing binaries with high-precision spaced-based photometry \citep{Prsa2011}. 

Several authors have investigated the eccentricity distribution of \textit{Kepler} binaries, which remains disputed. \citet{Vaneylen2016} reported a clear signature of tidal circularization and a significant difference between early- and late-type binaries. This result was since disputed by \citet{Kjurkchieva2017} and \citet{Windemuth2019} who reanalysed the \textit{Kepler} binaries and reported a larger fraction of eccentric binaries at short periods with no difference between early- and late-type systems. 

\begin{table}
\caption{Definition of model parameters}              
\label{table:params}      
\centering                                      
\begin{tabular}{c l}          
\hline\hline                        
Parameter & Definition \\    
\hline                                   
$t_1$ & Time of primary eclipse \\
$R_1$ & Stellar radius of primary \\
$T_1$ & Stellar effective temperature of primary \\
$a/R_1$ & Semi-major axis ratio \\
$P$ & Orbital period \\
$k$ & Radius ratio $R_2/R_1$ \\
$i$ & Orbital inclination \\
$e$ & Orbital eccentricity \\
$\omega$ & Longitude of periastron \\
$f_{\rm p}$ & Light ratio in TESS band \\
$f_{\rm p}'$ & Bolometric light ratio \\
$A_{\rm e}$ & Amplitude of ellipsoidal term $-A_{\rm e}\cos (4\pi t/P)$ \\
$A_{\rm r}$ & Amplitude of reflection term  $-A_{\rm r}\cos (2\pi t/P)$\\
$A_{\rm b}$ & Amplitude of beaming term  $A_{\rm b}\sin (2\pi t/P)$\\
$A_{\rm x}$ & Amplitude of phase shift term  $A_{\rm x}\sin (4\pi t/P)$\\
$f_0$ & Flux zero-point \\
\hline                                             
\end{tabular}
\end{table}
The most detailed studies are only possible for the brightest targets. The Transiting Exoplanet Survey Satellite \citep[TESS;][]{Ricker2015} is performing a two year near-all-sky survey. TESS observed the southern hemisphere from July 2018 to July 2019, covering nearly $130,000$ of the brightest stars in the southern sky in 2 minutes cadence over 13 sectors of 27 days duration. Near the ecliptic south pole $36,000$ targets were observed for at least two sectors. During it nominal 2~yr mission TESS will cover nearly the complete celestial sphere and thereby record for nearly all nearby, bright, short-period eclipsing binaries high-precision spaced-based photometry. This opens the possibility to characterize in detail these brightest short period eclipsing binaries in the sky, which should prove particularly insightful with respect to stellar physics.

Here we introduce a new catalog of eclipsing binaries located in the southern hemisphere based on TESS data from its first mission year. We describe the creation of our catalog in \S~\ref{sec:TESS_catalog}. The validation of orbital parameters is outlined in \S~\ref{sec:TESS_sample}. In \S~\ref{sec:combined_sample} we combine our catalog of TESS binaries with \textit{Kepler} binaries and binaries from the catalog compiled by \citet{Torres2010} and use this sample to investigate the orbital eccentricity distribution with respect to orbital separation and stellar effective temperature. We review the predictions of tidal theory in \S~\ref{sec:tidal_theory_predict} and use our new sample to test predictions from tidal circularization theories in  \S~\ref{sec:testing_tidal_theory}.
In \S~\ref{sec:alt_path_circ} we discuss alternative pathways to circularization and finally we summarise and conclude in \S~\ref{sec:conclusions}.

\section{Creating a Catalog of TESS Eclipsing Binaries} \label{sec:TESS_catalog}
To find and characterise eclipsing binaries, we performed a number of steps detailed below. We first searched all single-sector light curves for eclipses. We pruned the sample for instrumental and astrophysical false positives and visually inspected the list of possible binaries. To obtain orbital parameters of our eclipsing binary candidates, we model the eclipses of the TESS light curves. We then combine best-fit light curve solutions with the spectral energy distributions \citep[using broadband photometry and \textit{Gaia} DR2 parallaxes,][]{Gaia_2016, Gaia_2018} to obtain absolute stellar parameters of both components of the binaries. 

\subsection{Eclipse Search and Visual Inspection}
We have analysed a total of $247, 565$ single-sector short-cadence TESS light curves from sectors 1-13, covering TESS's first year of observing the ecliptic southern hemisphere. Starting with the Pre-search Data Conditioning SAP (PDCSAP) flux, we computed Lomb-Scargle periodograms of all light curves. If no significant power was detected in the power spectrum, we excluded the target from further analysis. This excluded $51\%$ of the targets. For the remaining $119, 219$ targets, we performed a transit search using the Transit Least Squares (TLS) algorithm developed by \cite{Hippke2019}. We ran the TLS search using the $\texttt{grazing}$-mode, which cross-correlates the light curve with a V-shaped model optimized for eclipsing binaries and grazing, transiting planets. $84\%$ of the targets had no significant signal in the transit search and were excluded from further analysis, leaving us with a list of $18, 531$ light curves. 

\begin{figure}
    \centering
    \includegraphics[width=\columnwidth]{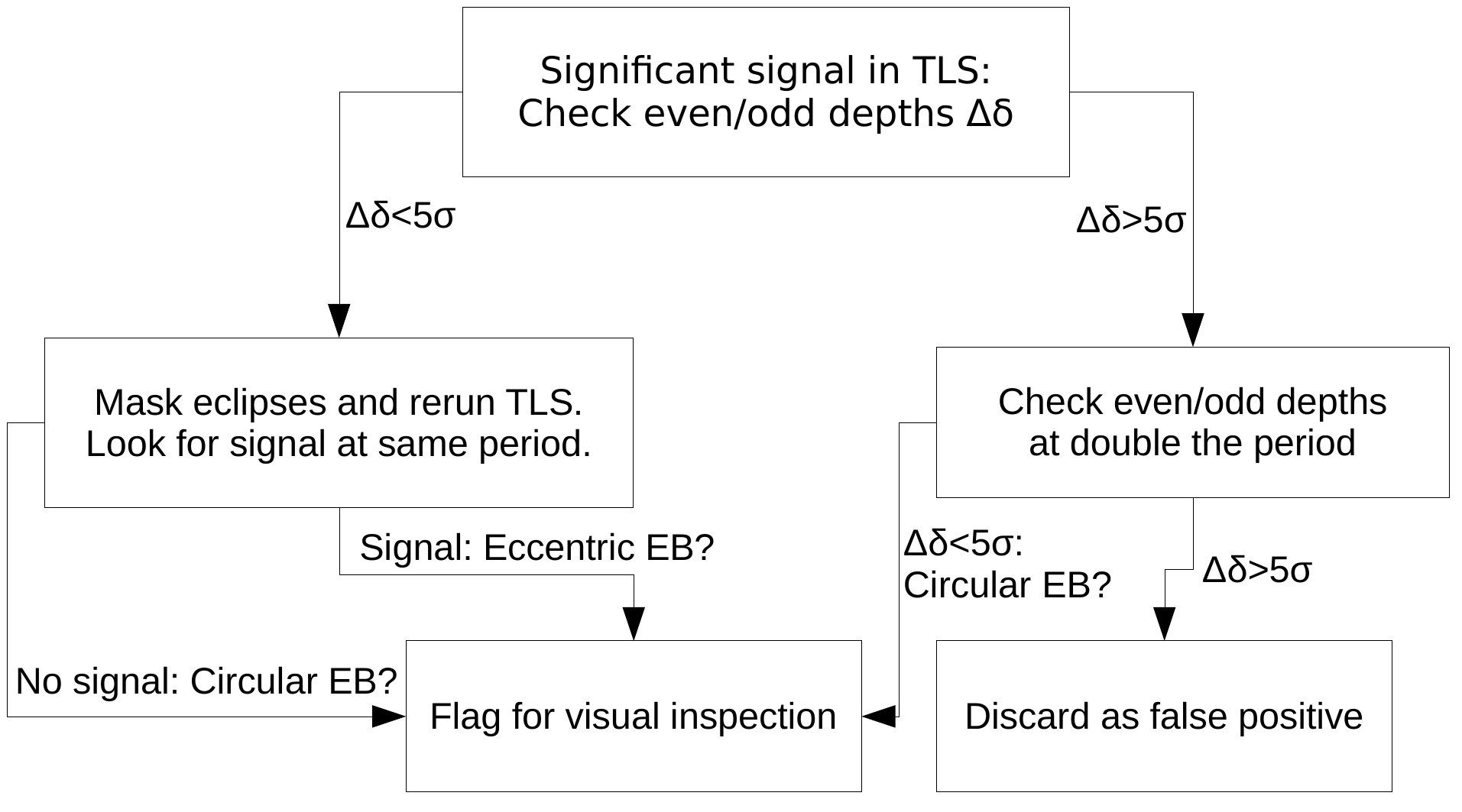}
    \caption{Flow chart illustrating the process of detecting circular and eccentric eclipsing binaries using even/odd eclipse depths.}
    \label{fig:flowchart}
\end{figure}

To ensure that we detect the correct orbital periods and to remove false positives, we compare the even/odd eclipse depths for each candidate system. We have visualised this process in Fig. \ref{fig:flowchart}.  If the even/odd depths agree within $5\sigma$, we likely found a circular equal-mass binary or the primary eclipses of an eccentric binary. 

The primary and secondary eclipses of eccentric binaries may not be separated by half an orbital period. To check for secondary eclipses, we therefore mask the primary eclipses and rerun the TLS algorithm to check for a second signal at the same period. If we detect a second set of eclipses within $3\sigma$ of the original period, we flag the target for visual inspection as an eccentric binary. If we do not detect a significant secondary signal, we flag the target for visual inspection as circular binary. If even/odd depths differ by more than $5\sigma$, we likely found a false positive or the half-period of a circular binary. In that case we check the consistency of the depths of every second eclipse. If the depths of every other eclipse agree within $5\sigma$, we flag the target for visual inspection as a circular binary. If the depths of every second event differ by more than $5\sigma$, we discard the target as a false positive. To limit the number of false positives, exoplanets and background binaries, we discard all targets with eclipse depths less than $1\%$. In total, $5,214$ light curves were flagged for visual inspection. 

During visual inspection, we discard any false positives that passed the even/odd test (exoplanets, pulsating stars, instrumental features, etc.). We also remove systems with shallow, low signal-to-noise (S/N) eclipses that passed through the initial tests but are likely background binaries and cannot be modelled reliably by our automated routine. We similarly discard binaries with light curves dominated by ellipsoidal variations or without clearly defined eclipses. This leaves us with a list of \nnotreject\ light curves with eclipsing binary candidates for which we have determined preliminary orbital periods, eclipse depths and primary and secondary eclipse timings. These systems advance to the final stage of orbital and stellar modelling.

\subsection{Deriving Orbital and Stellar Parameters} \label{ssec:deriving_parameters} 
We now begin our orbital and stellar modelling by preparing the light curves. We detrend the PDCSAP light curves after masking eclipses using a moving median filter with a window width equal to the orbital period. We do not attempt to remove features on shorter time scales than the orbital period due to the risk of affecting the eclipses. For systems that are observed in more than one TESS sector, we stitch their single-sector light curves before detrending. We iteratively remove $5\sigma$ positive outliers from the detrended light curves. 

With the light curve preparation complete we now model the eclipses using the quadratically limb-darkened models of \citet{MandelAlgol2002} as implemented in the Python package \texttt{batman} \citep{Kreidberg2015}. We model phase curve modulations using the BEER model \citep{Sirko2003, Mazeh2010}, a linear combination of four sinusoidal signals that describe the phase curve variation (see Table \ref{table:params}). It is necessary to model phase curve variations because we do not remove features on time scales shorter than the orbital period during detrending.

\subsubsection{Orbital Parameters} \label{ssec:orbital_params}
We model the light curves in several iterations. In the first iteration, we revise the orbital period and mid-eclipse timings from the TLS analysis by performing a least-squares fit for the orbital period ($P$) and phases of the primary ($t_1$) and secondary eclipses ($t_2$) while fixing other model parameters to their TLS values. In the second iteration we additionally include in the least-squares fit the radius ratio ($k$), scaled orbital separation ($a/R_1$), orbital inclination ($i$), two combinations ($\sqrt{e} \cos \omega$, $\sqrt{e} \sin \omega$) of the orbital eccentricity ($e$) and argument of periastron ($\omega$), the component light ratio in the TESS band ($f_{\rm p}$), and the BEER parameters ($A_{\rm e}$, $A_{\rm r}$, $A_{\rm b}$, $A_{\rm x}$, $f_0$). The PDCSAP flux is corrected for third light contamination by estimating the flux contribution of identified sources within 10 pixels of the target aperture \citep[for details see][]{KeplerHandbook2017, Stassun2019}. Since the PDCSAP flux is corrected for contamination, we do not fit a third light parameter. See Table \ref{table:params} for a full list of parameter definitions. We denote the deepest eclipse the primary eclipse. We use subscript $1$ for parameters of the primary component and subscript $2$ for the secondary component. We fix the limb darkening parameters to the values found by interpolating the \citet{Claret2017} limb darkening table for the TESS band using the stellar parameters listed in the TESS Input Catalog (TIC). Having optimized the orbital period $P$, we phase-fold the light curve on the orbital period and bin the light curve in two minute bins, using the median value of each bin. The median binning does not reduce the temporal resolution of the light curve but reduces the amount of data points and consequently increases computational speed in the next iteration. We create masks of 1.5 times the eclipse duration centered around the primary and secondary eclipse and refit our full model to the phase-folded, binned light curve using the MCMC sampler \texttt{emcee} \citep{ForemanMackey2013}. We run the MCMC sampler for $100,000$ iterations for each target{ using $48$ walkers initialized near the best-fit least-squares solution. We use the best-fit orbital parameters obtained in this MCMC analysis to derive stellar parameters as described in Sec. \ref{ssec:stellar_params}.}

\subsubsection{Stellar Parameters} \label{ssec:stellar_params}
Having obtained best-fit orbital parameters, we now go on to derive stellar parameters of the two components of the binaries. This is done by performing a joint fit to the TESS light curve and Spectral Energy Distribution (SED). We construct the SED from the apparent magnitudes, distance (adopted from \citet{2018AJ....156...58B} using \textit{Gaia} DR2 parallaxes) and extinction listed in the TIC. We include Johnson-Cousins $B$ and $V$, 2MASS $J$, $H$ and $K$ and \textit{WISE} $W1$ and $W2$ magnitudes where available. If no extinction is listed in the TIC, we query the 3D galactic dust map by \citet{Green2019} or the dust map by \citet{Schlegel1998} as the last priority. We convert apparent magnitudes to absolute magnitudes via the distance modulus. To obtain synthetic absolute magnitudes, we use the bolometric correction (BC) tables developed for the BaSTI isochrones \citep{Hidalgo2018} and compute the combined absolute magnitudes of the two binary components using their effective temperatures and radii\footnote{We adopt the BC tables assuming solar metallicity and a surface gravity of $\log g = 4.5$. We note that the BC is mostly insensitive to metallicity and surface gravity.}. We jointly fit the phase-folded, binned light curve and the SED while varying the radius ratio $(k)$, stellar effective temperature ratio $(T_2/T_1)$, stellar effective temperature of the primary component $(T_1)$ and stellar radius of the primary $(R_1)$ using $100, 000$ MCMC samples with \texttt{emcee}. The limb darkening parameters of both components are updated at each iteration to match their respective effective temperatures.  The light ratio in the wide red-optical TESS band $f_{\rm p}$ is different from the bolometric light ratio $f_{\rm p}' = (T_2/T_1)^4 k^2$. To ensure a consistent TESS light ratio, we therefore convert the bolometric light ratio to the TESS light ratio using the TESS BC at each iteration. The remaining orbital parameters are fixed to their best-fit values obtained in Sec. \ref{ssec:orbital_params}.

We have found that the SED generally provides poor constraints on the radius ratio and TESS light ratio. These values are therefore essentially fixed by the light curve and will change only slightly in the joint fit due to varying the limb darkening parameters. The SED does however allow us to derive absolute temperatures and radii that are consistent with the light curve solution. As a final step, we fit the light curve (now without the SED) while varying the orbital parameters $t_1$, $a/R_1$, $i$, $\sqrt{e} \cos \omega$, $\sqrt{e} \sin \omega$. This is done to accommodate any changes in these parameters as a response to the updated limb darkening parameters. For stars without distance measurements, we adopt the effective temperature listed in the TIC as the primary effective temperature and skip the SED fit.

\begin{figure}
    \includegraphics[width=\columnwidth]{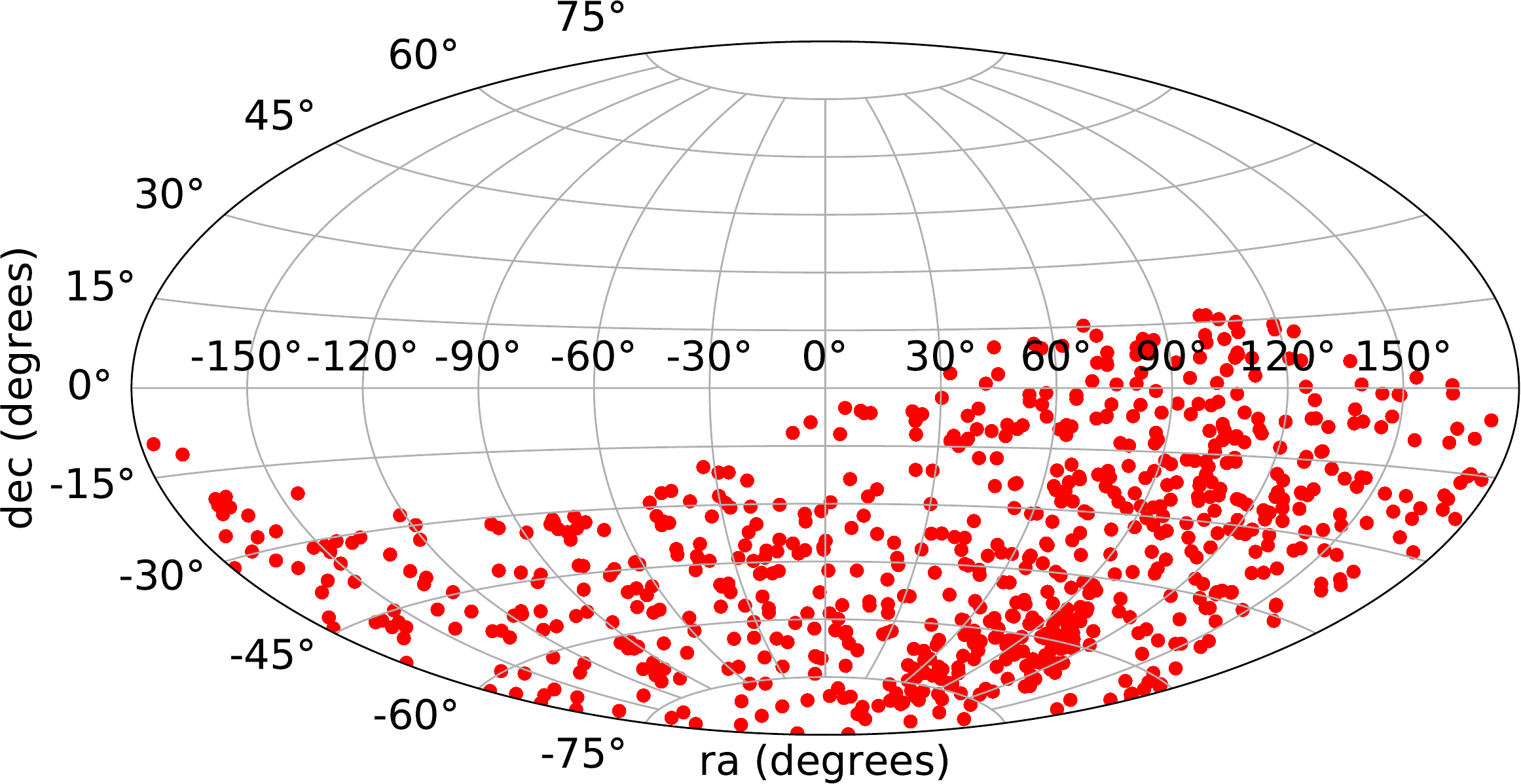}
    \caption{Distribution of detected TESS eclipsing binaries on the sky. The continuous viewing zone is visible as the over-density of binaries around $(\text{ra, dec}) \approx (90, -60)^{\circ}$.}
    \label{fig:TESS_sky}
\end{figure}

Finally, all \nnotreject\ systems are visually vetted a final time to ensure that the light curve and SED are modelled correctly. During this process, \nremovedduringfitting\ systems were removed due to poor light curve or SED modelling, leaving us with a catalog of \nfull\ eclipsing binary candidates. We show the sky-projected distribution of the TESS eclipsing binaries in Fig.~\ref{fig:TESS_sky}. The TESS Continuous Viewing Zone is visible as an over-density of binaries, indicating that binaries present in multiple sectors have a larger chance of being detected. We report orbital periods, time of primary and secondary eclipse, minimum eccentricity $\rm{abs(}e\cos\omega)$, eclipse depths and TESS magnitudes for all \nfull\ systems in Table \ref{table:orb_params}. In Table \ref{table:best_fit_params}, we list best-fit orbital and stellar parameters of a selection of \nreliable\ eclipsing binaries determined to have reliable parameters (see Sec. \ref{sec:tess_reliable_sample}). We do not report uncertainties on individual systems due to the difficulty in ensuring well-sampled posterior distributions in a highly degenerate parameter space. We run our MCMC chains for $100,000$ iterations (long enough to reach an acceptable best-fit solution) while a fully converged solution typically requires many millions of iterations \citep{Windemuth2019}. {To gauge the reliability of our derived parameters, we compare our TESS binary sample with independent results in Sec. \ref{ssec:validating_tess_cat}}. 

\section{The TESS Eclipsing Binary Sample} \label{sec:TESS_sample}
Having created a catalog of TESS eclipsing binaries, we now estimate the completeness of our catalog and validate our derived orbital and stellar parameters by comparing with other catalogs of eclipsing double stars. {In particular, we review the accuracy of the derived orbital eccentricities, which is crucial to robustly test tidal orbital circularization.}

\begin{deluxetable*}{cccccccc}
\tablewidth{0pt}
\tablecaption{Orbital parameters of TESS eclipsing binaries.}              
\label{table:orb_params}      
\tablehead{
\colhead{TIC} & \colhead{$P$ (d)} & \colhead{$t_{\rm 1}$ ($\rm{BJD} - 2457000$)} &\colhead{$t_{\rm 2}$ ($\rm{BJD} - 2457000$)}  & \colhead{$\rm{abs(}e\cos\omega)$} & \colhead{$\delta_1$} & \colhead{$\delta_2$} & \colhead{TESS Magnitude}
}
\startdata
286191384 & 3.612096 & 1572.97 & 1574.78 & 0.000135 & 0.27 & 0.27 & 9.74 \\
269852699 & 5.136474 & 1653.93 & 1651.37 & 0.000055 & 0.12 & 0.10 & 10.97 \\
153709888 & 4.332752 & 1438.63 & 1440.80 & 0.000003 & 0.07 & 0.06 & 11.47 \\
307488184 & 10.066890 & 1578.30 & 1573.36 & 0.014594 & 0.15 & 0.01 & 10.41 \\
98478039 & 0.668014 & 1491.82 & 1491.49 & 0.006004 & 0.04 & 0.01 & 13.93 \\
\enddata
\tablecomments{Table \ref{table:orb_params} is published in its entirety in the machine-readable format.  A portion is shown here for guidance regarding its form and content.}
\end{deluxetable*}

\begin{deluxetable*}{cccccccccccc}
\tablewidth{0pt}
\tablecaption{Best-fit orbital and stellar parameters of TESS eclipsing binaries.}              
\label{table:best_fit_params}      
\tablehead{
\colhead{TIC} & \colhead{$P$ (d)} & \colhead{$t_{\rm 1}$ ($\rm{BJD} - 2457000$)} & \colhead{$k$} & \colhead{$a/R_1$} & \colhead{$e\cos\omega$} & \colhead{$e\sin\omega$} & \colhead{$i$ ($^{\circ}$)} & \colhead{$f_{\rm p}$} & \colhead{$T_1$ (K)} & \colhead{$T_2$ (K)} & \colhead{$T_{\rm eff}$ from SED fit}
}
\startdata
286191384 & 3.612096 & 1572.97 & 0.99 & 8.35 & -0.000135 & -0.014339 & 84.80 & 0.96 & 7315 & 7250 & Yes \\
269852699 & 5.136474 & 1653.93 & 0.60 & 12.46 & 0.000055 & -0.038014 & 85.56 & 0.24 & 6172 & 5507 & Yes \\
153709888 & 4.332752 & 1438.63 & 0.42 & 9.12 & -0.000003 & -0.037847 & 83.91 & 0.13 & 6030 & 5517 & Yes \\
307488184 & 10.066890 & 1578.30 & 0.36 & 22.21 & 0.014594 & -0.047812 & 89.53 & 0.01 & 5610 & 3432 & Yes \\
185980914 & 3.491477 & 1519.53 & 1.20 & 5.44 & -0.000004 & -0.011463 & 88.09 & 1.21 & 7888 & 7424 & Yes \\
\enddata
\tablecomments{Table \ref{table:best_fit_params} is published in its entirety in the machine-readable format.  A portion is shown here for guidance regarding its form and content.}
\end{deluxetable*}

\subsection{Completeness of the TESS Catalog} \label{ssec:completeness}
To test the completeness of our catalog, we crossmatched the catalog of eclipsing binaries compiled by \citet{Malkov2006} with the list of targets observed by TESS in its first year. We selected all detached binaries (LCType \texttt{EA} in the \citet{Malkov2006} catalog) and found 176 targets observed by TESS in sector 1-13. 138 of these targets had passed our detection criteria and were flagged for visual inspection in our initial search.  We then visually inspected the light curves of the 38 targets not found in our initial search to understand why they were missing. Of these targets, 25 targets had either no or only one primary and/or secondary eclipse in their TESS light curves, 4 targets had very short periods {($P<0.5\,$d)}, 1 target had an eclipse depth less than $1\%$ and the remaining 8 targets were missed by our search due to large instrumental systematics or eclipses falling (partly) within gaps in the light curve. We note that even though we recovered $95\%$ of the \citet{Malkov2006} binaries within our search criteria, not all of these binaries appear in our final catalog, since they may have been removed during visual inspection due to low S/N, large ellipsoidal variations, poor best-fit solutions, etc. 

We emphasize that our catalog is not meant as an exhaustive catalog of eclipsing binaries in the TESS data. We have focused on detached binaries that can relatively easily be detected and modelled by our automatic routines. We have therefore rejected binaries not suitable for automatic modelling. \textit{Many} eclipsing binaries exist in the TESS data that are not included here, including thousands of semi-detached, contact and over contact binaries, ellipsoidal variables and a large number of shallow background binaries. We also do not include binaries without at least two sets of eclipses within a single sector. {This requirement limits the sample to binaries with orbital periods shorter than the $27\,$d duration of a \textit{TESS} sector.} An eclipsing binary may{ be excluded} in our catalog for a range of reasons: strong variability in the light curve (instrumental or stellar), too few eclipses in the light curve, shallow eclipses, low S/N eclipses, eclipses occurring near gaps in the light curve, light curve modelling fails to converge, etc. We exclude shallow or low S/N eclipsing binaries to avoid false positives and contamination from background binaries. To limit the number of false positives from the TLS search we check that the individually detected eclipses in the light curve have consistent depths. This requirement may remove some high S/N eclipsing binaries if the light curve is poorly detrended.

\subsection{The TESS Reliable Sample} \label{sec:tess_reliable_sample}
\begin{figure*}
    \centering
    \includegraphics[width=\textwidth]{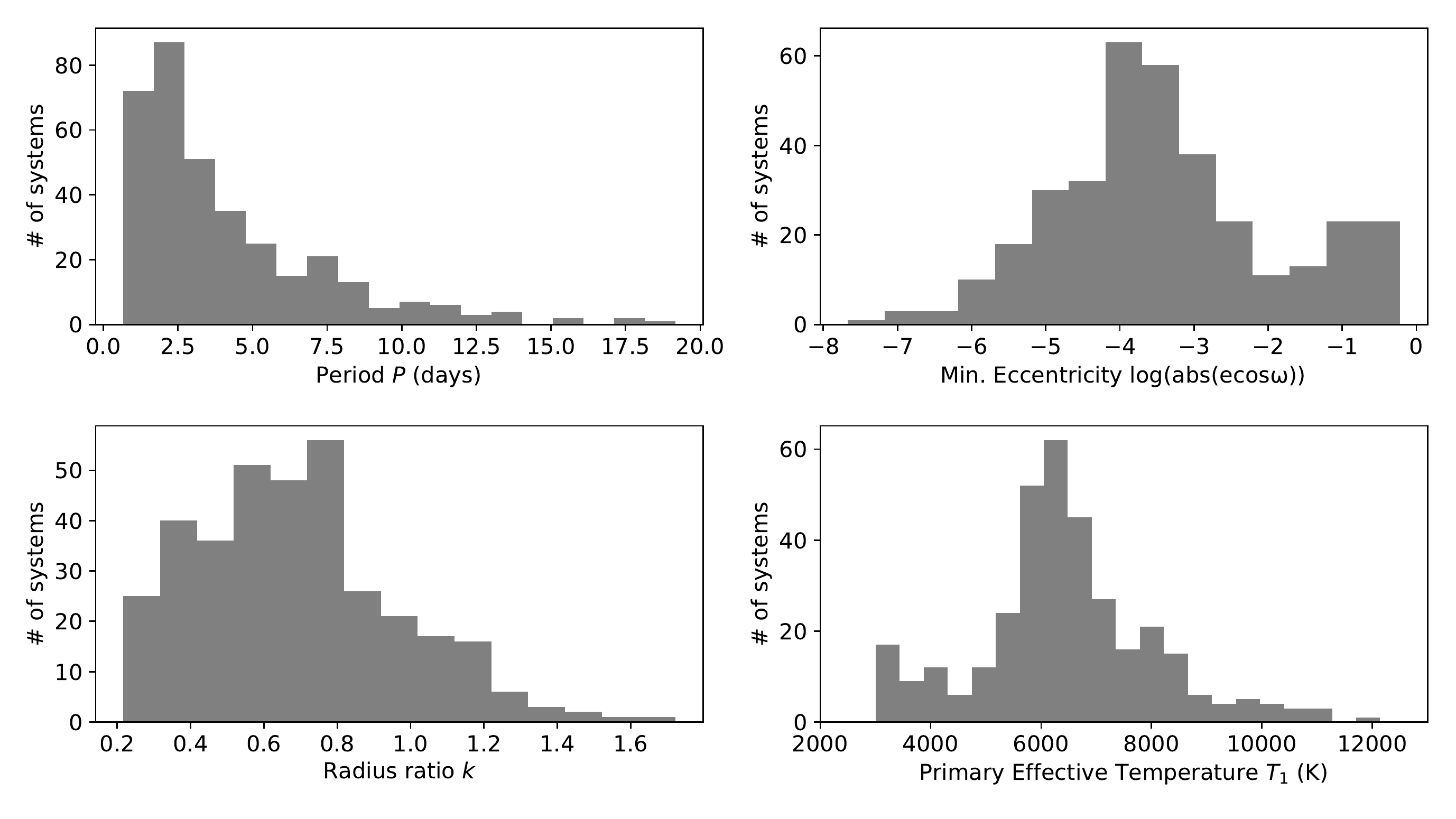}
    \caption{Orbital period, minimum eccentricity, radius ratio and primary effective temperature distributions of the TESS reliable sample.}
    \label{fig:TESS_distributions}
\end{figure*}
To obtain a sample of TESS binaries with the most reliable parameters, we define a subset of binaries with low contamination and low impact parameters which meet the criteria $b < 1$ and $\texttt{contratio} < 0.5$, where {$b=(a/R_1)\cos i$} is the impact parameter and \texttt{contratio} is the contamination ratio as defined in the TESS Input Catalog \citep{Stassun2018, Stassun2019}. We do this because for grazing binaries with impact parameters larger than one, it is not possible to uniquely constrain the orbital parameters. We limit the contamination ratio to avoid background binaries and to minimize the effect of contamination on the derived parameters {(although all light curves are corrected for the estimated contamination)}. To further minimize the number of background binaries, we limit the TESS sample to binaries with primary eclipse depths of at least $5\%$ and secondary eclipse depths of $1\%$.  This subset contains \nreliable\ binaries. We perform a comparison with other surveys to gauge the overall uncertainty of this sample (Section \ref{ssec:validating_tess_cat}). The uncertainty of any individual system will depend on the light curve morphology, the signal-to-noise ratio, the robustness of detrending and the goodness of fit.

In Fig.~\ref{fig:TESS_distributions} we plot the distribution of orbital periods, minimum eccentricities, radius ratios and temperatures for the TESS sample of reliable binaries. The period distribution peaks at $\sim 2\,$days with no binaries with periods above 20 days. This is a consequence of searching single sectors, which are 27 days long. The log-eccentricity distribution is double-peaked with a large fraction of $e\cos\omega \approx 0$ binaries and a smaller peak of binaries with minimum eccentricities in the range $0.1-1$. 

The binaries cover a wide range of temperatures from $2,000\,$K up to $50,000\,$K. For visual clarity we plot only the distribution of temperatures up to $13,000\,$K, {as only $5$ binaries in the reliable sample have hotter effective temperatures.} We stress that these distributions are strongly affected by selection biases and therefore do not represent the underlying distribution of eclipsing binaries in the sky. 

\subsection{Validating the TESS Reliable Sample} \label{ssec:validating_tess_cat}
Although many of the binaries in our catalog are known eclipsing binaries, a direct comparison with literature is complicated by the inhomogeneity of the available information for these binaries. We select here two works to validate i) our routines and ii) results. 

\begin{figure*}
    \centering
    \includegraphics[width=\columnwidth]{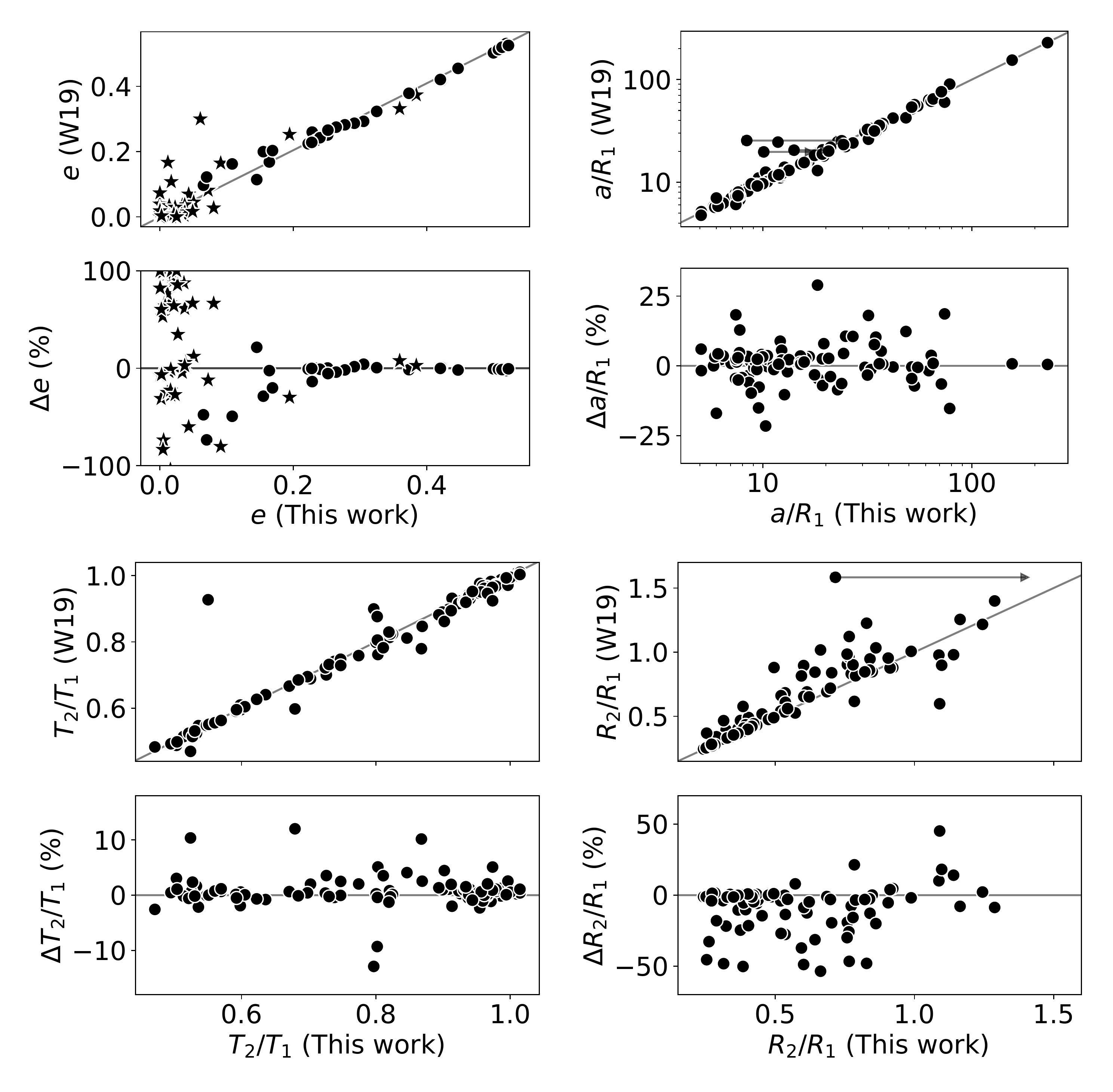}
    \includegraphics[width=\columnwidth]{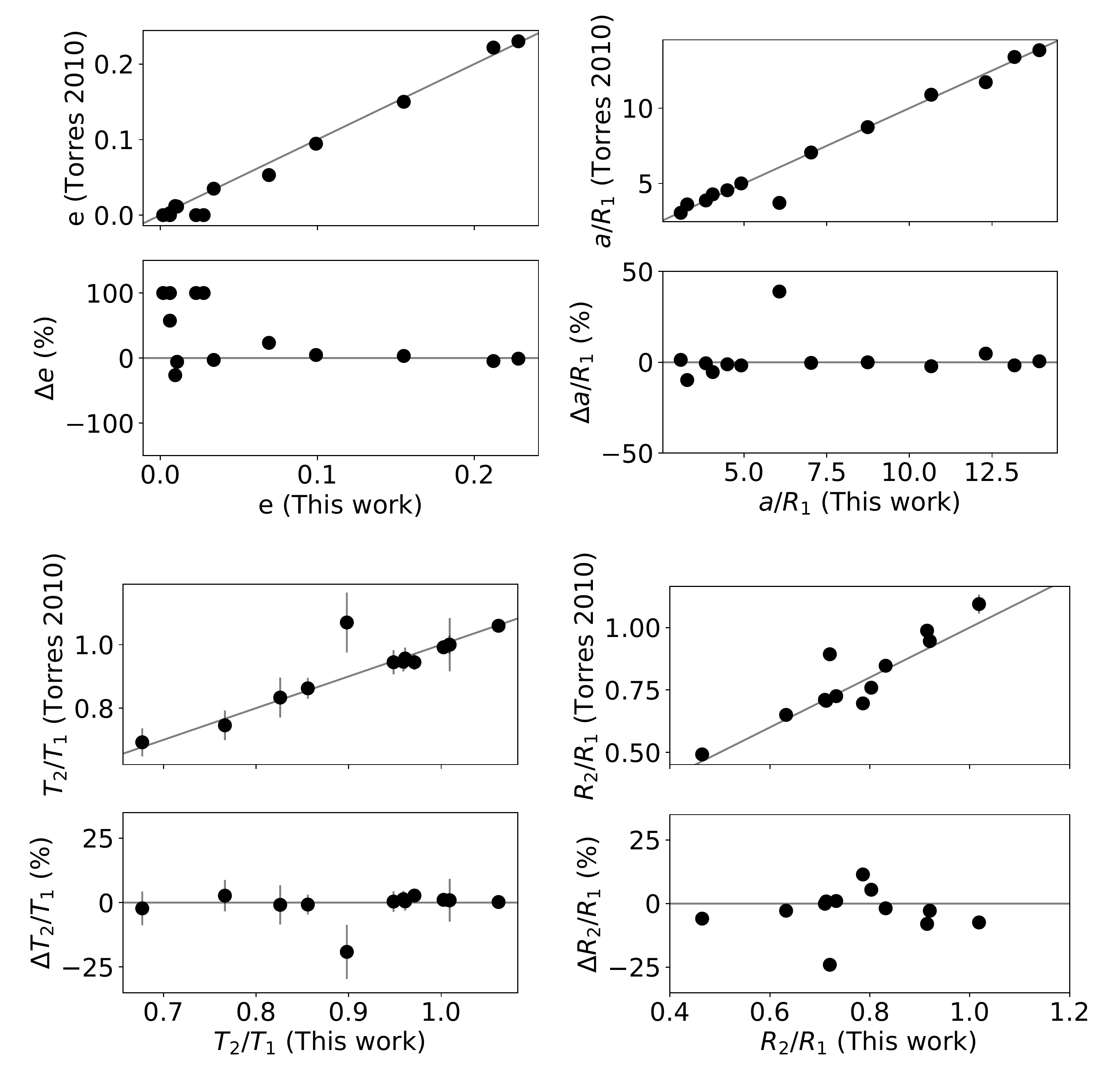}
    \caption{Comparison of eccentricity $e$, scaled semi-major axis $a/R_1$, temperature ratio  $T_2/T_1$ and radius ratio $k$ with 94 systems from \citet{Windemuth2019} (W19 in plots) and 13 systems from \citet{Torres2010}. \textit{Kepler} systems with $|e\cos\omega| < 0.05$ are marked as squares in the upper left panel. The best-fit parameters of \citet{Windemuth2019} do not have statistical uncertainties. The error bars on the \citet{Torres2010} temperature ratios are computed assuming uncorrelated errors of $T_1$ and $T_2$. The error bars on the remaining \citet{Torres2010} parameters are smaller than the symbol size for most data points.}
    \label{fig:Torres_and_Windemuth_compar}
\end{figure*}

The \textit{Kepler} eclipsing binaries (EBs) provide a homogeneous sample of well-studied eclipsing binaries with space-based, high-cadence, high precision light curves, similar to TESS. Recently, \citet{Windemuth2019} conducted a homogenous analysis of 728 \textit{Kepler} EBs by performing for each of these systems a joint fit of the light curve and the SED coupled to a grid of stellar isochrones, inferring consistent orbital and stellar parameters of both components in the binaries. Our southern hemisphere sample does not overlap with the \textit{Kepler} sample. We therefore analysed a randomly chosen subset of \textit{Kepler} binaries using our procedure developed for TESS EBs as described in Section \ref{sec:TESS_catalog}. This approach allows us to validate the accuracy of our analysis against the results of another team using a homogeneous sample of systems with excellent light curves. We compare systems with morphology parameters $\texttt{morph} < 0.5$ and stellar ages $\log \tau \text{/yr}> 7.5$, as recommended by \citet{Windemuth2019}. We restrict the comparison to match the TESS reliable sample with the criteria $b < 1$ and primary and secondary eclipse depths less than $5\%$ and $1\%$. We note that there are some important differences between the \textit{Kepler} and TESS samples: Most \textit{Kepler} binaries are observed in 30~minute cadence, the \textit{Kepler} light curves have much longer duration and generally much higher signal-to-noise S/N. Due to the longer duration, the median-binned light curve is less sensitive to poor detrending. The increased S/N makes it easier to visually identify poor fits. The \textit{Kepler} comparison sample is therefore potentially biased towards overall better accuracy than the TESS sample.

In Fig.~\ref{fig:Torres_and_Windemuth_compar} we compare eccentricities $e$, semi-major axis ratios $a/R_1$, temperature ratios $T_1/T_2$ and radius ratios $k$ between 94 \textit{Kepler} systems.
We find good agreement {(better than $10\%$ for most systems)} in semi-major axis ratios for all but three systems. The three outliers are the three systems with the largest radius ratios ($k > 1.5$) in \citet{Windemuth2019}. We find better agreement for these systems if the primary and secondary component are switched (as indicated by arrows in Fig.~\ref{fig:Torres_and_Windemuth_compar}). Temperature ratios show excellent agreement {(better than $5\%$)}. The relatively poor agreement {(within $50\%$)} in the radius ratios is expected due to the difficulty in uniquely determining radius ratios of grazing systems. Eccentricities show generally good agreement {(within $10\%$)} for values larger than $\sim 0.2$. We do however observe significant disagreement for the smallest eccentricities.

The derivation of reliable eccentricities $e$ and periastron angles $\omega$ from photometry is non-trivial. This is particularly true for circular or near-circular systems. 

The eccentricity vector $e\cos \omega$ is directly constrained via relative eclipse timings \citep[e.g.][]{Winn2010}
\begin{align}
    e\cos \omega \approx \frac{\pi}{2P}\left(t_2 - t_1\right) - \frac{\pi}{4}, \label{eq:ecosw}
\end{align}
where $t$ is the mid-eclipse time. While some corrections to this formula apply for high eccentricities, $e\cos\omega$ is essentially model-independent and can be determined reliably for all orbital configurations. The eccentricity vector $e\sin \omega$ is similarly constrained by the relative durations of the eclipses \citep[e.g.][]{Winn2010} 
\begin{align}
    e \sin \omega \approx \frac{T_2/T_1 - 1}{T_2/T_1 + 1}, \label{eq:esinw}
\end{align}
where $T$ is the eclipse duration from first to fourth contact point. However, while Eq. \eqref{eq:ecosw} is nearly always a good approximation, Eq. \eqref{eq:esinw} provides only a lower limit of $e\sin \omega$ and may not be accurate for grazing configurations \citep{deKort1954}. Furthermore, the determination of $e\cos\omega$ is significantly more precise than $e\sin \omega$ due to the difficulty in measuring precise eclipse durations, compared to determining eclipse timing.

For binaries with small eccentricities and equally spaced eclipses, the derived value of the eccentricity $e = [(e\cos\omega)^2 + (e\sin\omega)^2]^{1/2}$ is therefore dominated by the uncertainty in $e\sin\omega$. For a binary with $e\cos\omega \approx 0$, even a small $2\%$ difference in the fitted duration of the primary and secondary eclipse will lead to an eccentricity of $e=0.01$ (and $\omega \approx 90^{\circ}$ or $\omega \approx 270^{\circ}$), making it difficult to distinguish between systems with zero eccentricity and small but significant eccentricities.

The median difference \citep[this work $-$][]{Windemuth2019} of eccentricities is $\Delta e = -0.002^{+0.02}_{-0.04}$, while the median difference in $e\cos\omega$ is $2\cdot 10^{-4} \pm 5\cdot 10^{-4}$, confirming that $e\cos\omega$ is significantly more precise. For \textit{Kepler} binaries with $|e\cos\omega| < 0.001$, $70\%$ of the systems have a fitted value of $\omega$ within $3^{\circ}$ of $90^{\circ}$ or $270^{\circ}$, indicating that the derived eccentricities of these systems are likely overestimated due to slight differences in the fitted eclipse durations. We therefore caution against trusting the derived eccentricities of systems that have near-zero $e\cos\omega$ values and $\omega$ near $90^{\circ}$ or $270^{\circ}$.

The second sample we use for validation is presented in the work by \citet{Torres2010}.  \citet{Torres2010} compiled a list of eclipsing binaries with precise absolute stellar and orbital parameters. This sample has the advantage that we can compare directly the parameters derived in this work from TESS light curves with precise, independently derived parameters obtained via spectroscopic and photometric studies. In contrast to binaries in the \textit{Kepler} field, some binaries from the \citet{Torres2010} sample have been observed by TESS during its first year of observations. 18 binaries reported in \citet{Torres2010} were observed by TESS in sector 1-13. Three of these were discarded from our catalog for having only one set of eclipses (2 systems) or a poor best-fit solution (1 system). Two systems failed the criteria for the reliable sample by having a too large impact parameter (1 system) or too shallow eclipses (1 system). The remaining 13 systems are included in our TESS reliable sample. In Fig.~\ref{fig:Torres_and_Windemuth_compar} we compare the eccentricities, temperature ratios, radius ratios and semi-major axes ratios of the 13 binaries present in both samples. We find an overall excellent agreement between the two samples {(with typical agreement within $10\%$ in eccentricities larger than $0.05$, $5\%$ in temperature ratios, $10\%$ in radius ratios and $5\%$ in semi-major axis ratios)}. As expected, the radius ratios show the largest scatter.  {We also test absolute effective temperatures against the temperatures of \citet{Torres2010} and find that primary and secondary temperatures are accurate to $\sim\! 10\%$ (with median differences \citep[this work $-$][]{Torres2010} $\Delta T_1 = 2^{+7}_{-11}\%$ and $\Delta T_2 = -2\pm9\%$).} We note that the binaries in \citet{Torres2010} are generally bright, well-detached and non-grazing and therefore presents a best-case scenario for our analysis.

\section{The Eccentricity Distribution of Eclipsing Binaries} \label{sec:combined_sample}
Having established the good agreement between our TESS reliable sample with the works of \citet{Windemuth2019} and \citet{Torres2010}, we combine the three samples to obtain a large sample of eclipsing binaries suitable for investigating tidal circularization\footnote{{\bf Upon publication we will place interactive plots of our sample at} \url{phys.au.dk/exoplanets}.}. Before analysing the combined sample, we review the eccentricity distribution as derived from our TESS reliable sample and from \textit{Kepler} binaries. 

\subsection{The Eccentricity Distribution of TESS Binaries} \label{ssec:TESS_EBs_ecc_dist}

\begin{figure}
    \centering
    \includegraphics[width=\columnwidth]{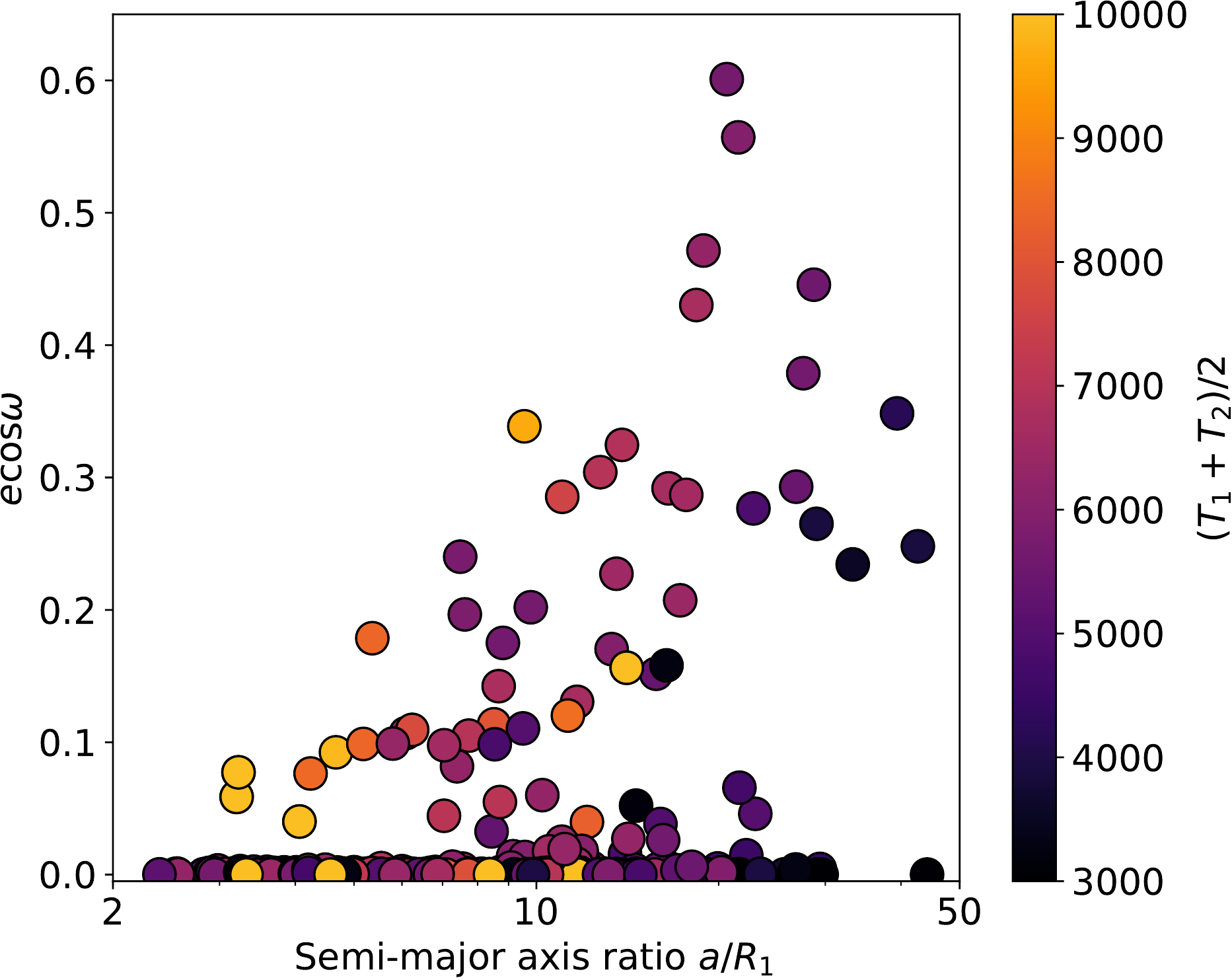}
    \caption{Eccentricity ($e\cos\omega$) distribution of the TESS reliable sample as function of semi-major axis ratio $a/R_1$. Systems are colorcoded corresponding to the mean effective temperature of the two components.}
    \label{fig:aR_ecc_TESS}
\end{figure}

In Fig. \ref{fig:aR_ecc_TESS} we plot the eccentricity distribution of the TESS reliable sample. Due to the difficulty in determining small eccentricities precisely from photometry, we use the much more precisely determined $e\cos\omega$ values as a proxy for eccentricity.\footnote{We note that $e\cos\omega$ may take both positive and negative values. Since the sign is irrelevant for the interpretation of $e\cos\omega$ as the minimum eccentricity, we always refer to $e\cos\omega$ as the absolute value in text and plots.} We plot $e\cos\omega$ as a function of scaled orbital distance and color-code the mean effective temperature of the binaries. {Fig. \ref{fig:aR_ecc_TESS} shows that the sample is clearly split} between a population of circular binaries with $e\cos\omega \approx 0$ and a population of binaries with significant eccentricities. {Circular binaries exist out to distances of} $a/R_1 \sim 50$. {The first eccentric systems appear} at $a/R_1 \sim 3$ with a clear temperature-dependence of eccentricity as function of scaled orbital distance. The closest-in eccentric binaries are all hot ($T_{\rm eff} \gtrsim 10,000\,$K) while cooler binaries begin to appear eccentric at larger scaled distances. We discuss these features in the context of tidal circularization in detail in Sec. \ref{sec:tidal_theory_predict} and \ref{sec:testing_tidal_theory}.

\subsection{The Eccentricity Distribution of \textit{Kepler} Binaries} \label{ssec:Kepler_EB_ecc_dist}
The \textit{Kepler} mission surveyed more than $150,000$ stars during its four year nominal mission. The \textit{Kepler} data are therefore one of the most important sources for the study of eclipsing binaries. The eccentricity distribution of \textit{Kepler} eclipsing binaries has been investigated by several authors. \citet{Vaneylen2016} measured $e\cos \omega$ of 945 \textit{Kepler} EBs and reported a difference in the eccentricity distributions of hot-hot and cool-cool binaries at short periods. They found that hot binaries appear eccentric at periods above $\approx 2$ days and that hot binaries are generally more likely to be eccentric than cool binaries at periods below 10 days. \citet{Vaneylen2016} used a boundary temperature of $6250\,$K to classify hot and cool binaries.  \citet{Kjurkchieva2017} analysed 529 \textit{Kepler} EBs and found only a weak correlation between period and eccentricity, with no difference between hot and cool binaries. \citet{Windemuth2019} found that even though eccentricities generally increase with orbital period, a significant fraction of both hot and cool binaries are eccentric even at the shortest periods. \citet{Windemuth2019} did not find a difference in the eccentricity distribution of hot and cool binaries.  

Contrary to \citet{Vaneylen2016}, \citet{Kjurkchieva2017} and \citet{Windemuth2019} derived eccentricities $e$ and periastron angles $\omega$ and analysed the resulting eccentricity distributions using their derived eccentricities $e$. We have found that their eccentricities are systematically overestimated for circular or near-circular systems due to the difficulty in determining precise eccentricities as discussed in Sec. \ref{ssec:validating_tess_cat}. The bias in $e$ for $e\cos\omega \approx 0$ systems results in period-eccentricity distributions that do not show the signature of tidal circularization as expected from tidal theory. This explains the discrepancy between the findings of \citet{Vaneylen2016} and later studies. 

\subsection{The Eccentricity Distribution of \citet{Torres2010} Binaries}
\citet{Torres2010} analysed the eccentricity distribution as a function of both orbital period and scaled distance. They split their sample at $7000\,$K to separate binaries with convective and radiative envelopes. They found that all binaries are circular below $1.5\,$days or $a/R_1 \approx 4$. They found that binaries with convective envelopes circularlise more easily and up to longer periods and radiative binaries.

\begin{figure*}[t]
    \centering
    \includegraphics[width=\textwidth]{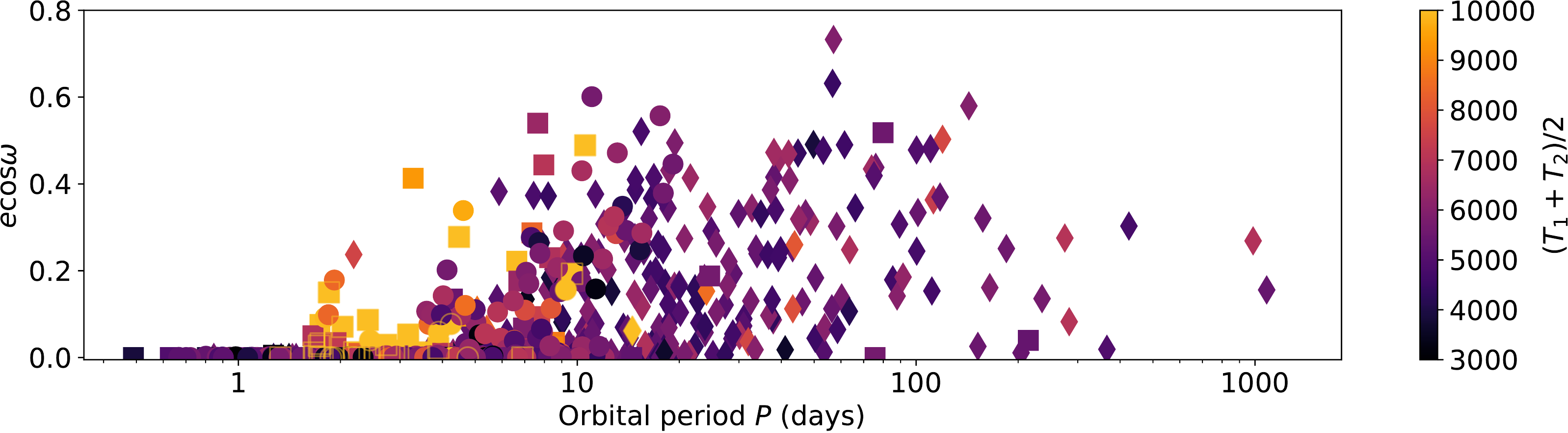}
    \includegraphics[width=\textwidth]{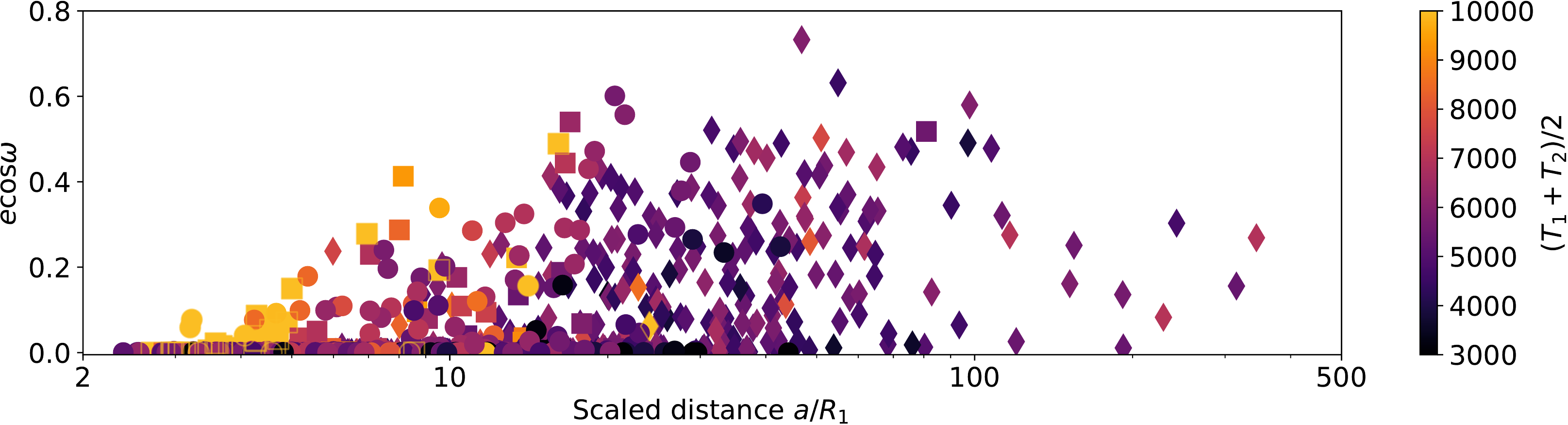}
    \caption{Eccentricity ($e\cos \omega$) distributions of the combined sample. \textit{Kepler} EBs from \citet{Windemuth2019} are marked with diamonds, \citet{Torres2010} EBs with squares and the TESS EBs with circles. Hot systems are plotted with open markers for visibility near $e \cos \omega \approx 0$. \textit{Upper panel:} $e\cos\omega$ as a function of orbital period. \textit{Lower panel:} $e\cos\omega$ as a function of scaled semi-major axis.
    }
    \label{fig:aR_ecc_combined}
\end{figure*}

\subsection{The Eccentricity Distribution of The Combined Sample}
We combine our TESS reliable sample with the \textit{Kepler} and \citet{Torres2010} samples to create a single data set. We do this to increase sample size and coverage in orbital separation. For systems common in our sample and in the catalog by \citet{Torres2010} we adopt the \citet{Torres2010} parameters. From the \textit{Kepler} sample we chose systems with $b < 1$, $\log\tau /\textrm{yr} > 7.5$ and $\texttt{morph} < 0.5$ as recommended by \citet{Windemuth2019}. As discussed previously, determining precise orbital eccentricities for systems with small values of $e\cos \omega$ using photometry alone is extremely challenging and prone to systematics. We therefore use the absolute value of $e\cos\omega$ as a proxy for eccentricity for \textit{Kepler} and TESS binaries. For binaries in \citet{Torres2010} we use the true eccentricities $e$ as these have been determined using a combination of photometry and radial velocities. The combined sample consists of \ncombined\ binaries, \ncombinedTESS\ from TESS, \ncombinedKEPLER\ from \textit{Kepler}, and \ncombinedTORRES\ systems from \citet{Torres2010}. This combined sample covers a range of $2-350$ in scaled semi-major axis $a/R_1$, ranges from $0.5$ up to $1000\,$days in orbital period and covers a temperature range of $3000-50,000\,$K in primary effective temperature.

The two panels in Fig.~\ref{fig:aR_ecc_combined} display the eccentricity distribution as a function of orbital period $P$ and scaled orbital distance $a/R_1$ for the combined binary sample. The (P, $e\cos\omega$) diagram is essentially model-independent while the ($a/R_1, e\cos \omega$) plot better represents the temperature-dependence of orbital eccentricity. As was observed in the TESS eccentricity distribution, we see a clear distance- and temperature-dependence of eccentricity. The first eccentric systems appear at $a/R_1 \approx 3.2$ or $P\approx 1.5\,$d. The two smallest separation eccentric binaries are HD\,152219 (TIC\,339566276) and $\delta$ Circini (TIC\,455509774). Both of these are O-type binaries with temperatures above $30,000\,$K. HD\,152219 is a member of the 2-7 million year old open cluster NGC\,6231 \citep{2006MNRAS.371...67S}. At $a/R_1 \sim 4-10$ we begin to see a larger population of hot binaries ($T_{\rm eff} \gtrsim 6000\,$K) with significant eccentricities. At $a/R_1 \gtrsim 10$, cooler binaries with convective envelopes begin to appear eccentric. We define a binary as eccentric when $e\cos\omega$ is larger than $0.005$ {(see Sec. \ref{ssec:pot_biases} for a discussion of this threshold)}. For $a/R_1 <10$, only $17\%$ of all systems are eccentric while for $a/R_1$ larger than $10$ $62\%$ of all systems are eccentric. We see a significant fraction of circular systems out to $a/R_1 \sim 50$. We note that the apparent decrease in maximum eccentricity beyond $a/R_1 \gtrsim 100$ or $P\gtrsim 200\,$d is most likely by chance due to a limited sample size and using the minimum eccentricity $e\cos\omega$ as a proxy of eccentricity.

\begin{figure*}
    \centering
    \includegraphics[width=\columnwidth]{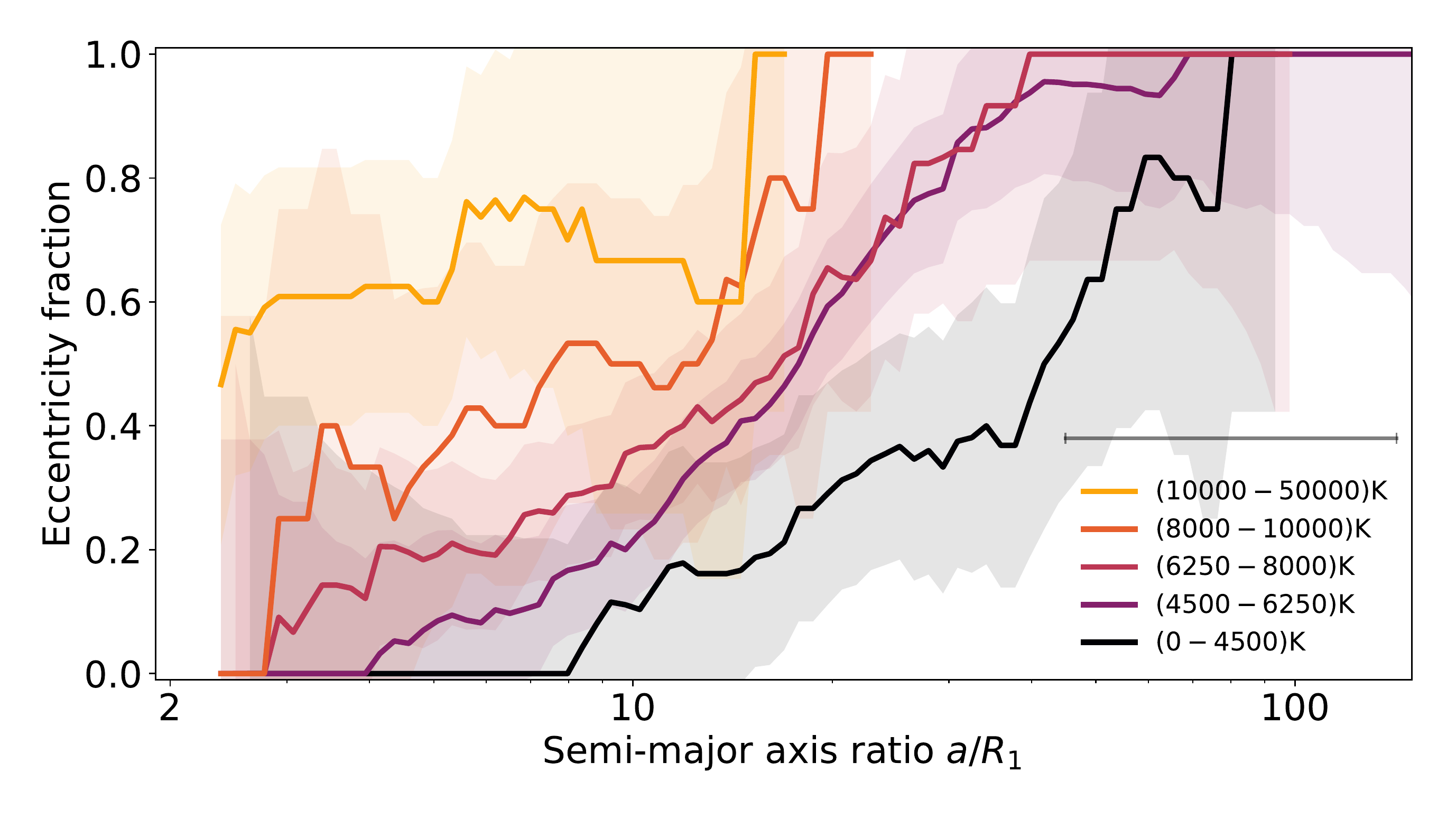}
    \includegraphics[width=\columnwidth]{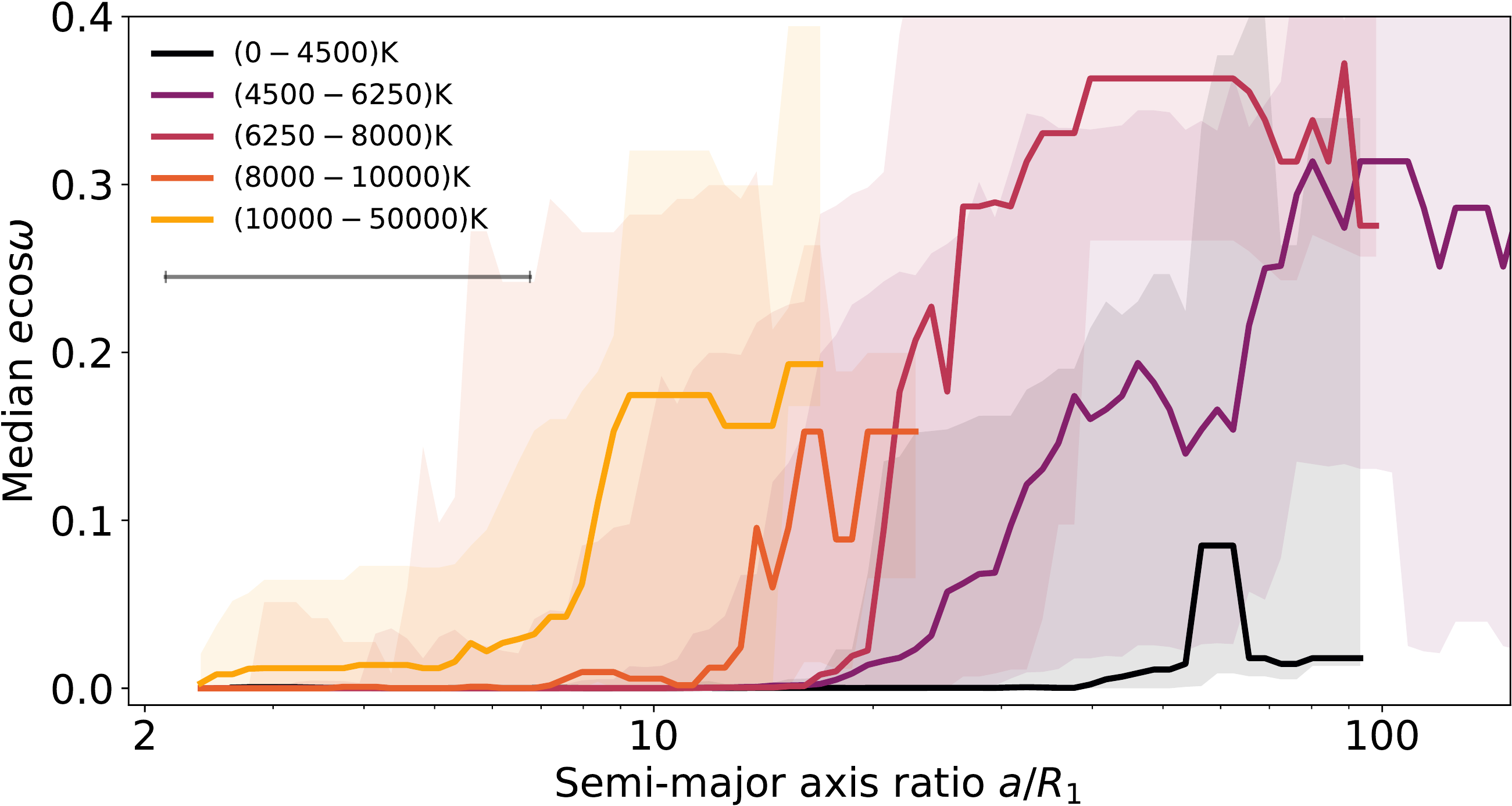}
    \caption{{\it Left panel}: Eccentricity fraction ($e\cos \omega > 0.005)$ as a function of semi-major axis ratio for five temperature bins in the combined sample. The shaded areas indicate uncertainties computed using Poisson statistics.  The grey bar represents the size of the sliding window. {\it Right panel}: Median eccentricity (using $e\cos\omega$ as a proxy for eccentricity for \textit{Kepler} and TESS binaries) as a function of semi-major axis ratio for five temperature bins in the combined sample. The shaded areas indicate the $68\%$ percentile regions ($i.e.$ an expected range of $0.16-0.84$ for a uniform eccentricity distribution). The grey bar represents the size of the sliding window.}
    \label{fig:combined_ecc_frac_and_median_ecc}
\end{figure*}

We present in Fig.~\ref{fig:combined_ecc_frac_and_median_ecc} the temperature-dependent eccentricity fraction and median eccentricity\footnote{Using true eccentricities $e$ for systems adopted from \citet{Torres2010} and $e\cos\omega$ for other systems} as a function of semi-major axis ratio. We have split the sample in five temperature bins. Only binaries in which the temperature of both components fall within the temperature intervals are included. This ensures roughly equal-mass binaries within the temperature intervals. The median eccentricity and eccentricity fractions are computed within a moving bin of width 0.5 in $\log a/R_1$-space and require at least 3 systems in a bin. {The eccentricity fraction is defined as $f(\log(a/R_1)) = N(e\cos\omega > 0.005)/N$ where $N\geq 3$ is the number of systems in the range $(\log(a/R_1) -0.25, \log(a/R_1) + 0.25)$.} The two panels of Fig.~\ref{fig:combined_ecc_frac_and_median_ecc} show that the orbital eccentricity correlates with the separation of the two components, as well as their temperature. The eccentricity fraction and median eccentricity increase with orbital separation and importantly with the effective temperature of the components. At all separations the eccentricity fraction and median eccentricity is higher in the higher temperature subsamples. The eccentricity fraction of all but the hottest binaries increases from zero to one. 

\subsection{Potential Biases} \label{ssec:pot_biases}
There are several potential biases in our combined sample. Using $e\cos\omega$ as a proxy for eccentricity introduces a risk of classifying an eccentric system as circular. For a population of equal-mass binaries with {uniform distributions} of $a/R_1$ in the range $2-150$, eccentricities $0-1$, periastron angles $0-360^{\circ}$ and impact parameters less than 1, $\sim\! 1\%$ of the eclipsing systems will have $e\cos\omega < 0.005$ and $e>0.005$. The combined \textit{Kepler} and TESS sample contains 420 systems with $e\cos\omega < 0.005$, of which 4 ($1\%$) have significant eclipse duration differences $T_1 - T_2 > 3\sigma$, in good agreement with the expected number of eccentric systems with low $e\cos\omega$. We therefore do not expect a significant fraction of eccentric binaries with very small $e\cos\omega$ values.

The division of circular and eccentric systems at $e\cos\omega = 0.005$ is somewhat arbitrary. The limit is however an order of magnitude larger than the estimated uncertainty of $e\cos\omega$ of $\sim 10^{-4}$. Our results are not sensitive to this limit. We have checked that the eccentricity fraction of the combined sample does not change significantly by halving the limit to $0.0025$ or doubling the limit to $0.0f1$.

The temperature intervals are chosen to divide our sample into stars with deep convection zones ($<4500\,$K), stars with convection zones ($4500-6250\,$K), stars with very shallow or no convection zones ($6250-8000\,$K), A-type stars without convection zones $(8000-10,000\,$K) and finally O- and B-type stars without convection zones ($>10,000\,$K). This division is chosen due to the different predictions from tidal theory for radiative and convective stars, as discussed in Sec. \ref{sec:tidal_theory_predict}. We have tested that our results are robust against varying temperature bin sizes and limits {by plotting the eccentricity fraction and median $e\cos\omega$ for a large number of temperature bins. We have similarly performed the analysis described in Sec. \ref{sec:testing_aRcrit} with bins of varying temperature limits and confirmed that the results are not sensitive to the exact limits of the temperature bins.}

It is easier to find circular binaries due to the higher combined S/N of equally spaced eclipses. This could bias our sample to an overall lower eccentricity fraction. However, since the combined sample is chosen to consist of well-modelled, high S/N light curves with deep eclipses, our sample is not limited by S/N. We therefore do not expect the eccentricity fraction to be affected by this detection bias. The eclipse probability is a function of eccentricity $e$ and periastron angle $\omega$ \citep[see e.g.][]{Winn2010}. For single eclipsing systems, the eclipse probability increases with eccentricity and peaks at $\omega = 90^{\circ}$, causing a bias in the derived eccentricity distribution if not corrected for \citep{Kipping2014}. This bias is not present in eclipsing binaries in which both the primary and secondary eclipse are visible. Due to the opposite dependencies of $\omega$ of the primary and secondary eclipse, the eclipse probability is essentially constant for all eccentricities and periastron angles (with the exception of extreme eccentricities $e\sim 1$). The eccentricity distribution is therefore not significantly affected by the eclipse probability.

\section{Predictions of Tidal Theory} \label{sec:tidal_theory_predict}
Having obtained the eccentricity distribution of a large sample of eclipsing binary stars with well-determined orbital parameters, we now proceed and use this new sample to test tidal theories. We specifically focus on the widely adopted theories of tidal dissipation in binary stars developed by \citet{Zahn1975, Zahn1977, Zahn1989a}. Zahn suggests two main dissipation mechanisms: the equilibrium tide which works for stars like our Sun and cooler stars with outer convective envelopes and the dynamical tide which works for hot stars with radiative envelopes. Below we will briefly summarize both mechanisms and their predictions. {We also highlight a few comparisons between theoretical predictions and observations.} For additional reviews of tidal theory and comparison to binary star observations see e.g. the works by \cite{Zahn2008} and \cite{Mazeh2008}.

\subsection{Circularization of Cool Stars} \label{ssec:cool_stars_circ}
\begin{figure}
    \centering
    \includegraphics[width=\columnwidth]{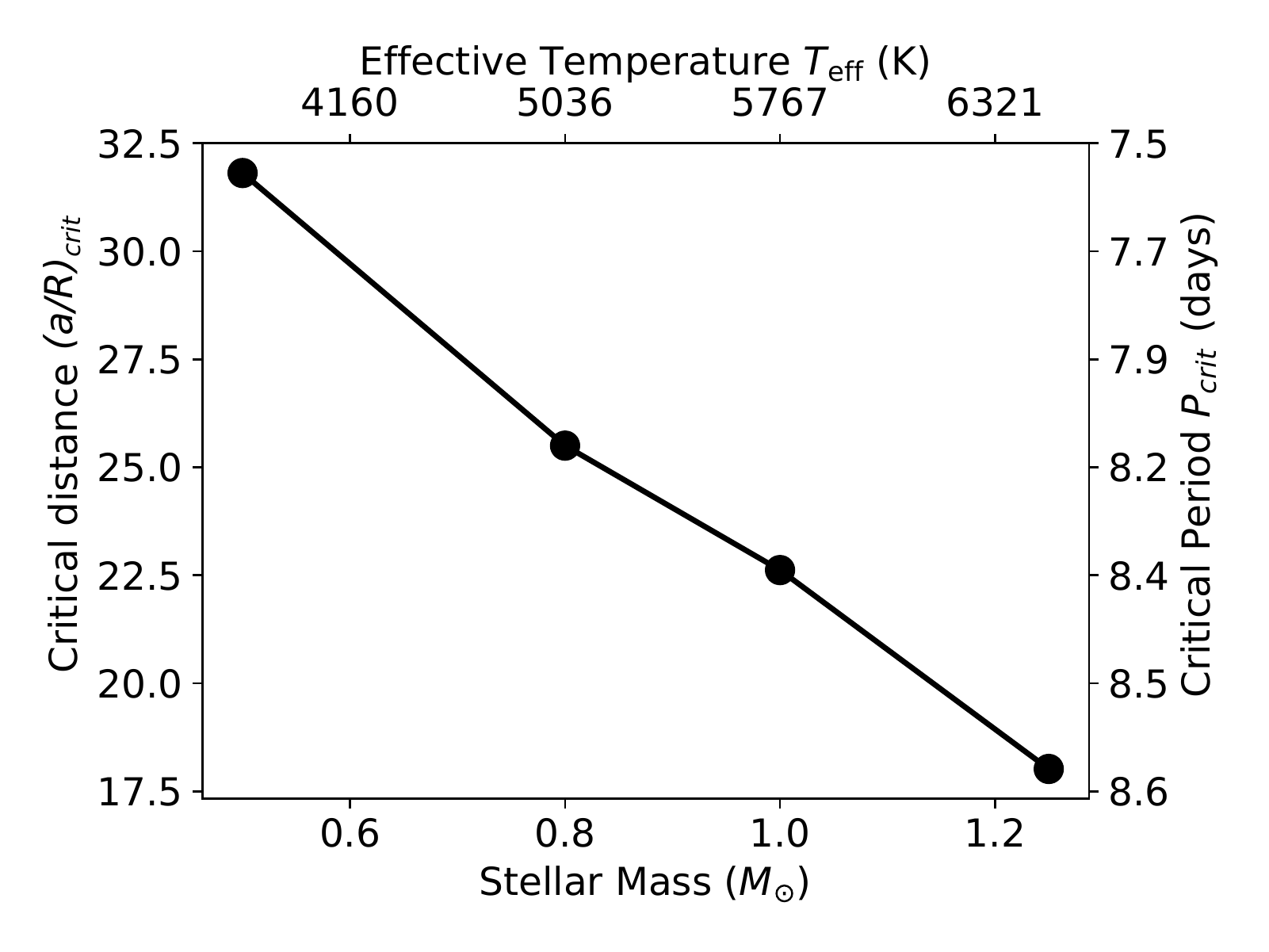}
    \caption{Critical circularization distance $(a/R)_{\rm crit}$ and period $P_{\rm crit}$ as a function of stellar mass and temperature for equal-mass binaries with convective envelopes in the revised equilibrium tide. Binaries are predicted to have circular orbits below these values. The critical periods are adopted from \citet[][Table 2]{Zahn1989b}. Critical distances have been computed from the critical periods using Kepler's third law and solar-metallicity BaSTI stellar models \citep{Hidalgo2018}.}
    \label{fig:equilibrium_tide_aR_Pc_M_Teff_ZahnBouchet1989_Table2}
\end{figure}
The dissipation mechanism of the equilibrium tide is turbulent friction in the convective envelope. This dissipation mechanism is more effective than the dynamical tide (discussed below), leading to the prediction that cool stars are circularised at larger orbital separations than hotter stars. It was originally assumed that tidal circularization occurs on the main sequence (MS).

\citet{Zahn1989a} and \citet{Zahn1989b} revised the equilibrium tidal theory by improving the treatment of convection within the mixing-length formalism and found that due to the larger size and extended convective envelope of cool late-type stars during the pre-MS (PMS), the equilibrium tide circularises binaries efficiently during this phase, with negligible subsequent circularization on the MS. 

The revised equilibrium tidal theory predicts that binaries with masses in the range $0.5 - 1.25 M_{\odot}$ are circularised out to periods of $7-9$ days independent of age \citet{Zahn1989b}, corresponding to scaled orbital distances of $18 - 32$ depending on stellar mass. We plot the critical circularization distance $(a/R)_{\rm crit}$ and period $P_{\rm crit}$ as a function of effective temperature and stellar mass in Figure~\ref{fig:equilibrium_tide_aR_Pc_M_Teff_ZahnBouchet1989_Table2}. The critical circularization distance $(a/R)_{\rm crit}$ is the distance which separates circularised and eccentric binaries. Binaries with orbital distances $a/R_1$ larger than their corresponding critical distances have circularization time scales longer than their stellar life times and are therefore not predicted to be circularised. 

The prediction that circularization occurs predominantly on the PMS has been challenged by measurements of tidal circularization in clusters of different ages \citep[e.g.][]{Mathieu1992, Claret1997, Mathieu2004, Meibom2005}. \citet{Meibom2005} measured the circularization period of eight late-type binary populations and found evidence that the circularization period increases gradually with age.{They found that the predicted circularization period of $\sim \! 7$ days agrees well with measurements of PMS binaries. However, the measured circularization period increases with age up to $\sim \! 16$ days for $10$ Gyr old halo binaries, in disagreement with the prediction of PMS circularization. }This indicates that while the equilibrium tide does circularise binaries up to 7 days during the PMS, there may be significant additional circularization throughout the main-sequence that cannot be explained by the revised equilibrium tide. 

\subsection{Circularization of Hot Stars} \label{ssec:hot_stars_circ}
\begin{figure}
    \centering
    \includegraphics[width=\columnwidth]{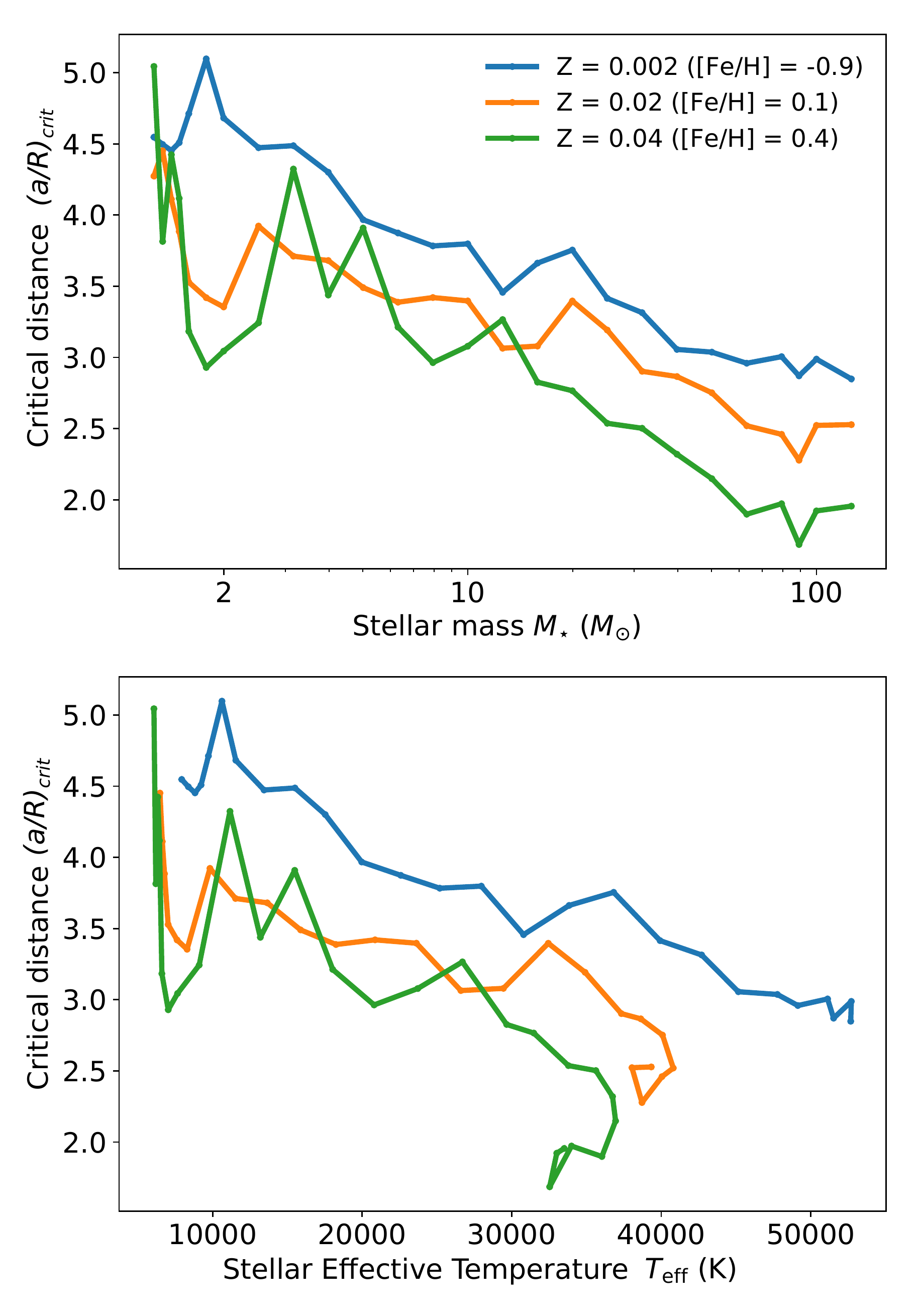}
    \caption{The critical circularization distance $(a/R)_{\rm crit}$ as a function of stellar mass (\textit{upper panel}) and stellar effective temperature (\textit{lower panel}) for {equal-mass binaries} at three metallicities {(see legend in upper panel)}. The critical circularization distance is computed in the framework of the dynamical tide using Eq. \eqref{eq:aRcrit}.}
    \label{fig:dyn_tide_mass_Teff_aR}
\end{figure}

Stars that lack an outer convective envelope cannot circularise via the equilibrium tide. The disappearance of the convective envelope is a continuous process occurring near the Kraft break at $\sim 6250\,$K (or $\sim 1.2\,M_{\odot}$) \citep{1967ApJ...150..551K}. 
For these stars, the dynamical tide is the dominant dissipation mechanism. The dynamical tide works by dampening of gravity modes in the radiative envelope. This dissipation mechanism is much less efficient than the equilibrium tide and occurs during the main-sequence.

\citet{Zahn1977} found the dynamical tide is extremely sensitive to $a/R_1$ with a time scale for orbital circularization of $t_{\rm circ} \propto (a/R_1)^{21/2}$. Due to the steep dependence of $a/R_1$, the tidal time scale is longer than the stellar life time for all but the closest binaries. For equal-mass binaries with synchronized spin and radiative envelopes, the critical circularization distance may be expressed as \citep{North2003, Zahn1977}
\begin{equation}
    \left(\frac{a}{R}\right)_{\rm crit} \approx 12.8 \cdot \left[\left(\frac{M_{\star}}{R^3 }\right)^{1/2} \int \left(\frac{R}{R_{\star}}\right)^9 E_2 d\tau \right]^{2/21} \label{eq:aRcrit}
\end{equation}
where $E_2$ is the tidal torque constant, $M_{\star}$ and $R_{\star}$ the stellar mass and radius in solar units and $\tau$ the stellar age in Myr. The tidal torque constant depends sensitively on the size of the convective core. Despite its name, the tidal torque constant is therefore a function of both stellar mass and age.

{Here, we estimate the critical circularization distance using modern stellar evolutionary tracks.} Using the evolutionary tracks by \citet{Claret2004, Claret2005, Claret2007} and Eq. \eqref{eq:aRcrit}, we compute the critical distance for equal-mass binaries in the range $1.25-125M_{\odot}$ for four metallicities, taking into account the evolution of stellar radius and tidal torque constant during the main-sequence. We display the results in  Fig.~\ref{fig:dyn_tide_mass_Teff_aR}, where we further assumed an age of half the main-sequence age. The result is not sensitive to the choice of stellar age. 

We find that while $(a/R)_{\rm crit}$ is a decreasing function of mass, it lies in the range $(a/R)_{\rm crit} \sim 3-4$ for a wide range of stellar masses. This agrees with the findings of \citet{North2003} who investigated the critical distance $(a/R)_{\rm crit}$ and found that it is insensitive to stellar mass, surface gravity and metallicity for stars of masses $5-15\,M_{\odot}$. \citet{North2003} found that the increase of $E_2$ with stellar mass cancels out the effect of decreasing stellar density and average age with mass and results in a critical distance of $a/R_1 \approx 4$ to within a few percent. They used zero-age main-sequences values of $E_2$  and neglected the change in stellar radius and convective core size during the first half of the main-sequence (replacing the integral in Eq.~\ref{eq:aRcrit} with $E_2 \tau$). As seen in Fig. \ref{fig:dyn_tide_mass_Teff_aR}, {we find that while including the effects of the change in stellar radius and convective core size results in only a slight decrease in critical distance from $(a/R)_{\rm crit}=3.5$ at $5\,M_{\odot}$ to $(a/R)_{\rm crit}=3.1$ at $15\,M_{\odot}$, the critical distance decreases by nearly a factor of two from $(a/R)_{\rm crit}= 4.25$ at $1.25\,M_\odot$ to $(a/R)_{\rm crit}=2.5$ at $100M_\odot$ (at [Fe/H]$=0.1\,$dex).} The critical distance depends sensitively on both the stellar radius and the convective core size and is therefore model-dependent. In addition, the tidal circularization time scale is derived under several simplifications and assumptions such as neglecting the effect of rotation and using a simplified description of convection \citep[see][]{Zahn1977}. 

{To compare the predicted critical circularization distance with observations, }\citet{North2003} investigated the circularization of B-type eclipsing binaries in the Magellanic Clouds (at sub-solar metallicity) using $e\cos \omega$ measurements from OGLE. {They found that} all binaries appear circular below a{ scaled orbital }distance of $a/R_1 = 3.85-4.1$, in good agreement with the prediction from the tidal theory. However, while no eccentric binaries are observed below this distance, a fraction of circular binaries were observed above the critical distance. This result was confirmed by \citet{Mazeh2006} who used a significantly larger sample of homogeneously analysed OGLE binaries and recovered the critical distance of $a/R_1 \approx 4$, while also observing a fraction of binaries with low eccentricities above the critical distance.

\begin{figure*}
    \centering
    \includegraphics[width=\columnwidth]{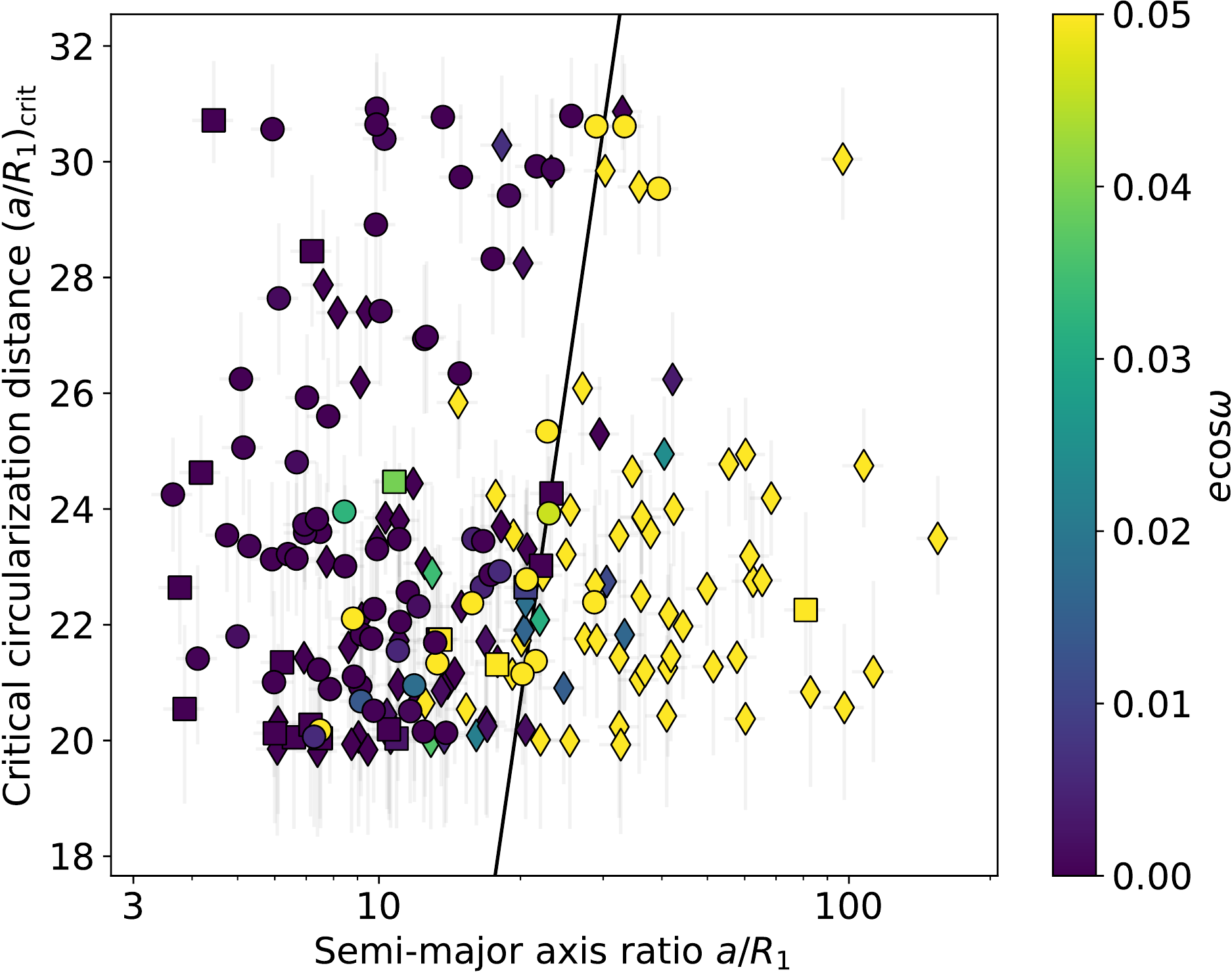}
    \includegraphics[width=\columnwidth]{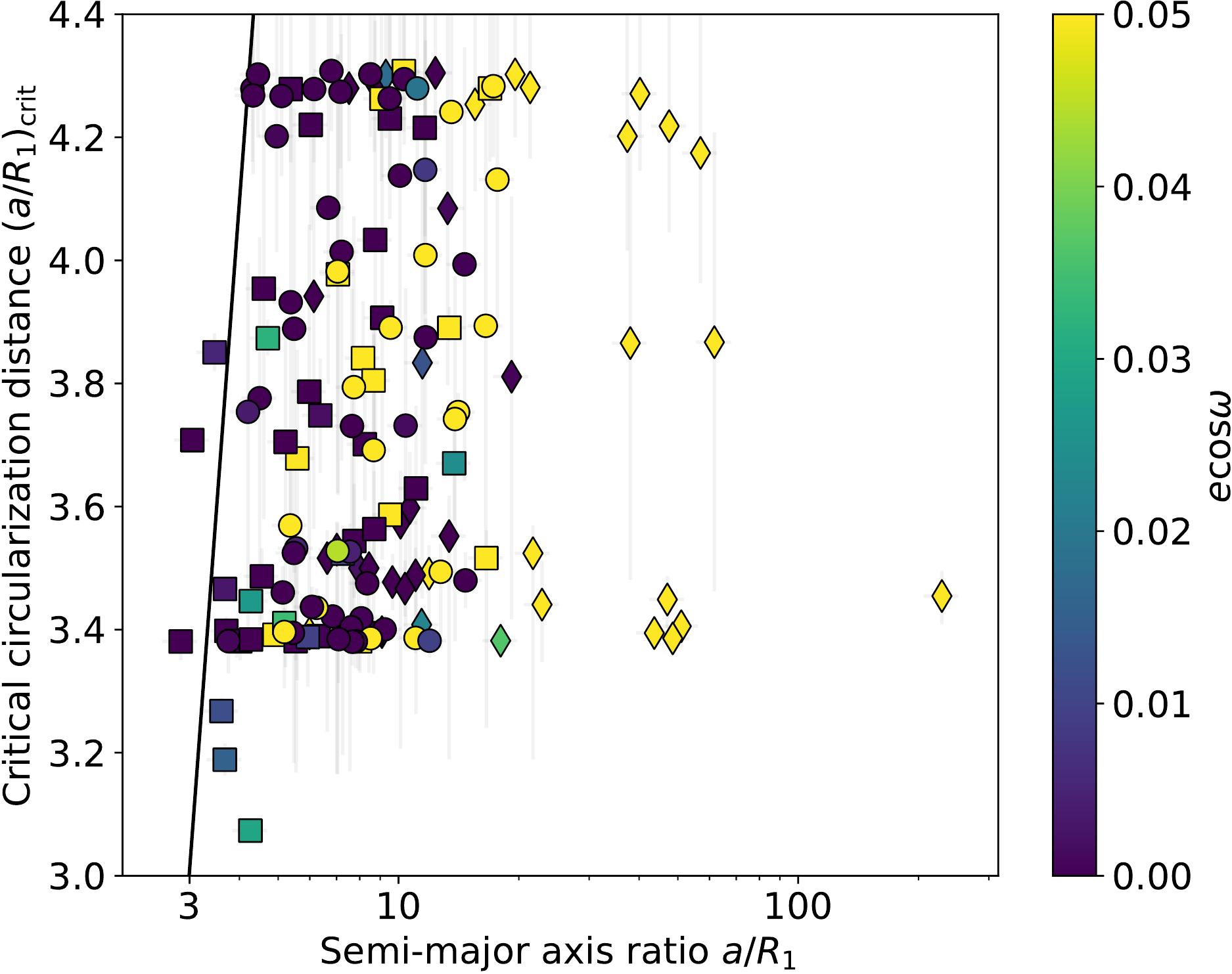}
    \caption{Critical circularization distance $(a/R)_{\rm crit}$ as function of scaled distance $a/R_1$. Error bars are computed by assuming a $250\,$K uncertainty on the effective temperature and a $10\,\%$ fractional uncertainty on $a/R_1$ of all targets. The vertical black lines represent the $a/R_1 = (a/R)_{\rm crit}$ line. Binaries to the left have semi-major axes smaller than the critical circularization distance and are expected to be circularised. \textit{Left panel:} Binaries with temperatures below $6250\,$K. The critical circularization distance is computed using the equilibrium tide of \citet{Zahn1989b}. \textit{Right panel:} Binaries with temperatures above $6329\,$K. The critical circularization distance is computed using the dynamical tide of \citet{Zahn1977} as described in Section \ref{ssec:hot_stars_circ}.} 
    \label{fig:aR_aRcrit_cool_and_hot}
\end{figure*}

\section{Testing Tidal Theory} \label{sec:testing_tidal_theory}
In a first test we compute the critical circularization distance $(a/R)_{\rm crit}$ for each system in the combined sample. Systems for which $a/R_1$ is smaller than $(a/R)_{\rm crit}$ should be circularised. In a second test we use the formalism introduced by \citet{Meibom2005} to robustly measure the circularization distance and period for each temperature interval. We summarize the results of these tests in Table~\ref{table:aR1_min_and_max}. {For both tests, we limit our sample to binaries with approximately equal-mass components. We do this in order to directly compare our observations with the predictions from the equilibrium and dynamical tide derived in Section \ref{sec:tidal_theory_predict}.}

\subsection{The Critical Circularization Distance, $(a/R)_{\rm crit}$} \label{sec:testing_aRcrit}
For binaries cooler than $6250\,$K we compute critical circularization distances in the equilibrium tide by interpolating the critical periods listed in \citet[][Table 2]{Zahn1989b}. We compute the critical circularization distance as a function of primary stellar effective temperature by converting the critical period to a scaled distance using Kepler's third law and solar-metallicity BaSTI stellar models \citep{Hidalgo2018}. For binaries hotter than $6329\,$K (corresponding to $1.25\,M_{\odot})$  we compute the circularization distance in the dynamical tide using Eq.~\eqref{eq:aRcrit} with the $Z = 0.02$ stellar models of \citet{Claret2004} as described in Section~\ref{ssec:hot_stars_circ}.

The critical distance depends on assumptions about the accuracy of stellar models, the logarithmic decrease in eccentricity (\textit{i.e.} the initial eccentricity), the initial radius on the PMS, the initial rotational velocity, the efficiency of tidal breaking, the mixing-length description of convection, neglecting differential rotation, neglecting magnetic activity, neglecting resonances, neglecting effects of additional companions, neglecting enhanced tidal forces of evolved stars and more \citep[see][]{Zahn1975, Zahn1977, Zahn1989a, Zahn2008, Zahn1989b}. Furthermore, we ignore the effect of metallicity in the stellar structure (using solar- or near-solar-metallicity stellar models) and assume an age of half the main-sequence lifetime for all hot stars (while convective stars are assumed to be circularised on the PMS). We likewise ignore the spread in companion masses and compute circularization distances under the assumption of equal-mass components.

Figure~\ref{fig:aR_aRcrit_cool_and_hot} shows the critical circularization distance as a function of semi-major axis ratio for binaries with components within $1000\,$K of each other in the combined sample. {We exclude systems in which the components differ by more than $1000\,$K to justify the assumption of equal mass.} The left panel shows cool stars, and the right panel shows the results for stars hotter than $6329$~K. 

Despite the above mentioned assumptions and limitations, we find for cool stars overall good agreement between observations and predictions made using the equilibrium tide formalism. We observe the majority of binaries with {scaled orbital} distances smaller than their critical distance to be circular as predicted. A small fraction of binaries are eccentric below their critical distance{ ($15$ of $145$ binaries have $a/R_1 < (a/R_1)_{\rm crit}$ and $e\cos\omega > 0.05$, although $8$ of these $15$ binaries have scaled distances within $10\%$ of their critical distances and only $2$ of $73$ binaries with $a/R_1 \leq 0.5(a/R_1)_{\rm crit}$ remain eccentric)}. {The eccentricity of these systems may be caused by the presence of tertiary companions \citep[][see also Sec. \ref{ssec:companions}]{2001ApJ...562.1012E, Tokovinin2006}.} {A tertiary companion may perturb the eccentricity of the close pair on a time scale similar to the circularization time scale. However, eccentricity modulation by a third companion may occur even when the eccentricity of the close-in pair is zero, suggesting that tertiary companions may be responsible for small, non-zero eccentricities (see \citet{Mazeh2008} and references therein).} Similarly, a small fraction of binaries are circular or near-circular at distances larger than their critical distances. We do not find evidence that these binaries are evolved \citep[which could explain increased tidal dissipation, see e.g.][]{sun_orbital_2018}. Their small eccentricities may represent the low eccentricity tail of the primordial eccentricity distribution.

For binaries hotter than $6329\,$K we do not see such good agreement between theory and observations. While we do see the first eccentric systems near the predicted critical distance of $(a/R)_{\rm crit} \sim 3$, a large fraction of systems are circular at distances larger than their critical distance predicted by theory. The majority of binaries are circular out to distances $\sim \! 3$ times larger than their predicted circularization distances. Assuming that these systems are not born circular this suggests that hot binaries are circularised significantly more efficiently than expected by the dynamical tide model. The disagreement between theory and observation is most profound below $10,000\,$~K. Above $10,000\,$K, only two out of 30 systems are circular beyond an $a/R_1$ of $5$.  Systems near the boundary of $6250-6329\,$K are sensitive to errors in their effective temperatures. If some binaries are cooler than expected, they may have circularised more efficiently via the equilibrium tide. However, we see increased tidal circularization also in the hotter $8000-10,000\,$K temperature bin (see Table \ref{table:aR1_min_and_max}). It is therefore unlikely that uncertain effective temperatures can explain the large number of hot circular binaries above the critical distance. 

\begin{table}
\caption{The number of systems $N$, fitted circularization distance $(a/R_1)_{\rm circ}^{\rm fit}$, theoretical critical distance $(a/R)_{\rm crit}$, fitted circularization period $P_{\rm circ}^{\rm fit}$ and theoretical critical period $P_{\rm crit}$ for each temperature bin in the combined binary sample.}
\label{table:aR1_min_and_max}
\centering
\begin{tabular}{llllll}
\hline\hline 
Temperature (K) & $N$ & $(a/R_1)_{\rm circ}^{\rm fit}$ & $(a/R)_{\rm crit}$ & $P_{\rm circ}^{\rm fit}$ (d) & $P_{\rm crit}$ (d) \\
\hline 
$0-4500$        & 58  & $29.8^{+42}_{-4}$     & $28 - 32$ & $5.57^{+0.20}_{-0.66}$ & $7-8$   \\
$4500-6250$     & 193 & $18.0^{+3.1}_{-0.86}$ & $19 - 28$ & $9.0^{+5.5}_{-2.0}$    & $7-9$   \\
$6250-8000$     & 109 & $11.4^{+3.5}_{-0.48}$ & $3 - 4$   & $5.5^{+3.3}_{-0.29}$   & $1-2$ \\
$8000-10000$   & 20  & $8.4^{+5.0}_{-1.9}$   & $3 - 4$   & $5.7^{+2.4}_{-1.9}$    & $1-2$   \\
$10000-50000$ & 29  & $3.8^{+4.7}_{-0.3}$   & $2 - 4$   & $2.8^{+0.84}_{-0.13}$  & $1-2$   \\
\end{tabular}
\end{table}

\subsection{The Circularization Function}

\begin{figure}
    \centering
    \includegraphics[width=\columnwidth]{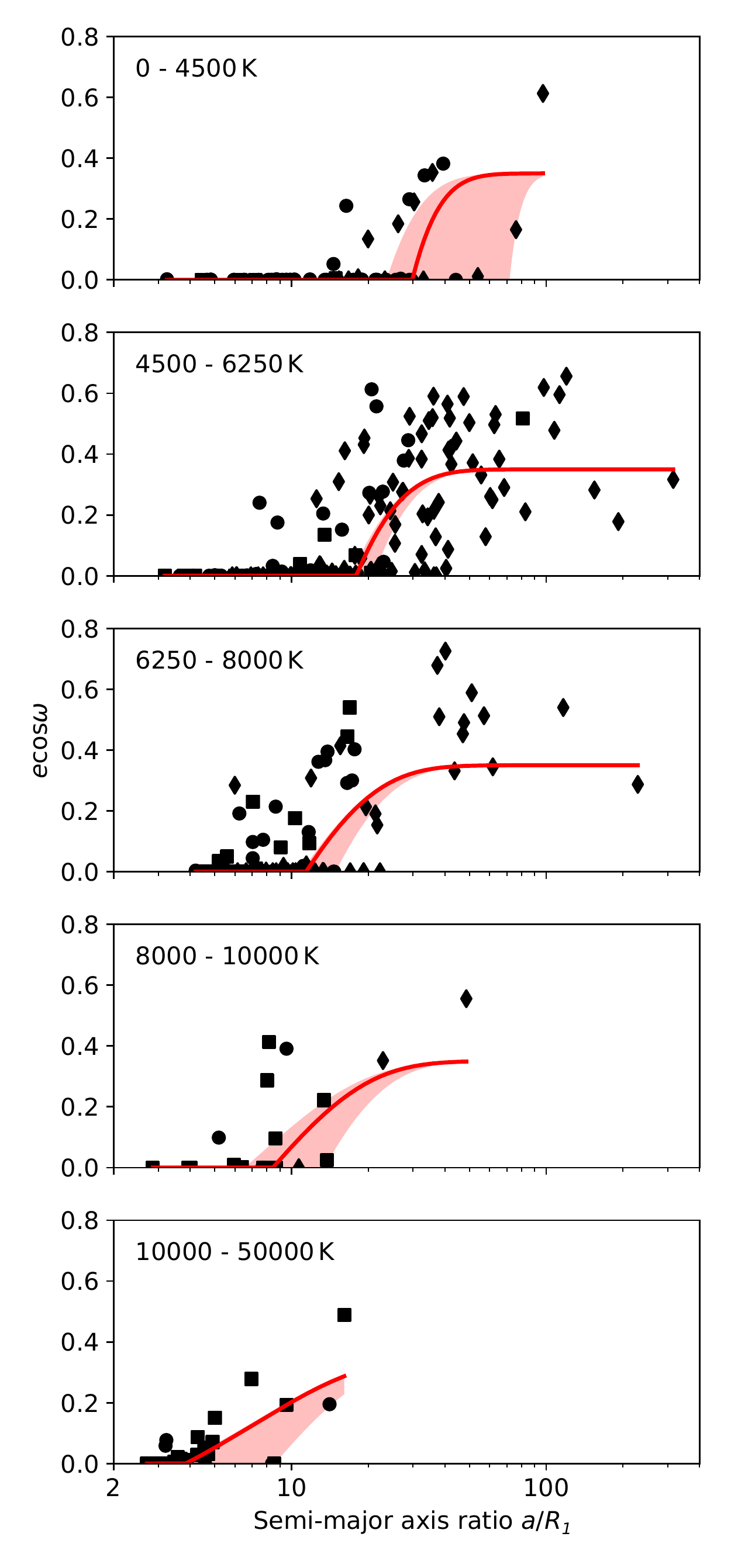}
    \caption{Eccentricity distribution of the combined sample in different temperature bins with the median best-fit circularization function in red. The $68\%$ uncertainty interval from bootstrapping is plotted in shaded red.}
    \label{fig:aR1_ecc_Meibomfits}
\end{figure}

As discussed by \citet{Meibom2005} and \citet{Mazeh2008}, the furthest-out circular binaries might represent the low-eccentricity tail of the initial eccentricity distribution and will therefore not represent the most likely circularization distance. To avoid the issue of simply comparing the furthest-out circular binaries to the critical circularization distance expected from theory, \citet{Meibom2005} introduced the circularization function $e(P)$ to robustly measure orbital circularization as a function of orbital period for a sample of binary systems. The function (which is not physically derived but motivated by numerical modelling of the eccentricity distribution) has the following form:
\begin{align}
e(P)=\begin{cases}
       0.0 \hfill \text{if $P \leq P'$, }\\
       \alpha(1-e^{\beta(P' - P)})^{\gamma}\,\,\,\,\,\,\,\,\, \hfill \text{if $P > P'$.}
     \end{cases}
\end{align}
We adopt the coefficients $\alpha = 0.35$, $\beta = 0.14$ and $\gamma = 1.0$ and fit the period $P'$ by minimizing the total absolute deviation as suggested by \citet{Meibom2005}. We extend the concept of the circularization function to {scaled orbital} distances ($a/R_1$) and have fitted the circularization function to the five temperature bins of the combined sample using both the orbital period $P$ and the semi-major axis ratio $a/R_1$ as the independent variable, see Table \ref{table:aR1_min_and_max}. {To minimize the influence of unequal-mass binaries, we include only binaries in which the effective temperature of both components fall within the same temperature bin.}

We discuss here the circularization function in the context of scaled distance. {Following \citet{Meibom2005},} the circularization distance $(a/R_1)_{\rm circ}$ is defined as the distance where the circularization function is equal to $0.01$. We determined the uncertainties on the fitted circularization distances via bootstrapping. We sampled each temperature bin $10,000$ times by redrawing all $N$ systems within the bin, where we allowed repeated draws. We list median and $84\%$ percentiles of the resulting best-fit solutions along with the expected range of critical circularization distances and periods from theory in Table \ref{table:aR1_min_and_max}.

We plot the {$e\cos \omega$ }distributions and circularization functions as a function of $a/R_1$ {for each temperature bin} in Figure~\ref{fig:aR1_ecc_Meibomfits}. { For systems with components cooler than $6250\,$K, the derived circularization distances are consistent with the theoretical critical distances of the equilibrium tide}. For stars in the temperature intervals $6250-8000\,$K and $8000-10,000\,$K we find circularization distances significantly larger than the expected critical distances, see Table \ref{table:aR1_min_and_max}. From the dynamical tide, we expect binaries hotter than $6250\,$K to circularise only below $a/R_1\sim 3-4$. We see that this is not the case. A significant number of binaries are found with circular orbits at significantly larger orbital separations{, even after accounting for the initial eccentricity distribution.} The circularization distance of binaries hotter than $10,000\,$K is however consistent with theory. This indicates an enhanced tidal dissipation mechanism for stars with temperatures $\sim 6000-10,000\,$K. These results all match the direct comparison with theory as seen in Fig. \ref{fig:aR_aRcrit_cool_and_hot} and discussed in Sec. \ref{sec:testing_aRcrit}.

\section{Alternative Pathways to Circular Orbits} \label{sec:alt_path_circ}
Although we do observe some cool binaries with low eccentricities at large orbital distances, we find that Zahn's theory of the equilibrium tide generally explains the observed eccentricity distribution of cool binaries. Our derived circularization period of $4500-62500\,$K binaries is in excellent agreement with theory. However, the $1\sigma$ upper limit of $P_{\rm circ}^{\rm upper} = 14.5\,$d (see Table \ref{table:aR1_min_and_max}) is consistent with the $\sim\!16\,$d circularization period of old Halo stars found by \citet{Meibom2005}. The large circularization period of old Halo stars reported by \citet{Meibom2005} has been seen as evidence of enhanced tidal dissipation of cool stars on the main-sequence. 

While cool binaries generally agree with theory, we observe a large fraction of circular hot binaries with large orbital distances which cannot be explained by the dynamical tide {(89 of 167 ($53\%$) of binaries with $T_{\rm eff,1,2}>6250\,$K and $a/R_1>4$ have $e\cos\omega < 0.005$)}. In this section we review possible mechanisms which could explain these observations.   

\subsection{Multiple Formation Pathways}
Recent hydrodynamic simulations have shown that there are many pathways to forming binary stars \citep{bate_formation_2002, bate_statistical_2019}. However, binary stars are generally thought to form on eccentric orbits. This assumption is supported by the increasing eccentricity fraction as a function of increasing orbital distance. We find that the eccentricity fraction reaches $100\%$ for all binaries at $a/R_1 \gtrsim 50$, consistent with the assumption that these binaries are unaffected by tidal forces. While it is possible that a small fraction of binaries form on primordially circular or near-circular orbits, this is unlikely to explain the large fraction of circular, hot binaries at orbital distances of $a/R_1\sim 4-10$.

\subsection{Complicated Orbital Evolution History}
Even among the hottest binaries we find circular systems at large orbital separations. The furthest-out circular binary in our $10,000+\,$K temperature interval is CV Velorum at $a/R_1 = 8.6$. CV Velorum is a $40\,$Myr old B-type ($T_{1, 2} \sim 18,000\,$K) binary. According to tidal theory, CV Velorum should not have been able to circularise at this distance. By observing the spectroscopic eclipse and analysing the Rossiter-McLaughlin effect, \citet{Albrecht2014} found that the rotation axes of both stars in the CV Velorum system are misaligned with respect to the orbital angular momentum and that the rotation of both components are consistent with synchronisation. Tidal theory predicts that synchronisation happens before circularization and that spin-orbit alignment occurs on a similar time scale as synchronisation. It is therefore surprising to find CV Velorum in a misaligned, circular and synchronized orbit at a distance where tides are not predicted to be significant. CV Velorum is not known to have close companions. The spin-orbit misaligned but circular configuration of CV Velorum might hint at a more complex formation and migration history than assumed in the standard applications of tidal theory{, e.g. involving interactions with a circumbinary disc or close encounters of nearby stellar systems early in its formation \citep[][]{bate_formation_2002}.} The existence of binaries with far-out circular orbits might therefore point towards a population of binaries with complex histories not well-described by the simplifying assumptions of tidal theory, one of which is the assumption of spin-obit alignment.

\subsection{Enhanced Tidal Dissipation}
If some binaries experience significantly enhanced tidal dissipation during their main-sequence evolution, this could explain the existence of circular binaries at large orbital distances. \citet{Witte1999, 2001A&A...366..840W} found that \textit{resonance locking} has the potential to greatly increase the efficiency of tidal dampening of massive stars on the main-sequence. They found that tidal dissipation is greatly increased during the main-sequence when the binary undergoes several periods of resonance locking of the tidal potential and stellar oscillation modes. They argue that resonance locking is a common phenomenon in the tidal evolution of eccentric{, hot} binaries. 

\citet{Witte2002} extended the study of resonance locking to stars with convective envelopes. They found that because the dynamical tide works in the radiative cores of convective stars on the main-sequence, resonance locking may also speed up the tidal evolution of solar-like stars. They investigated if this could explain the large circularization periods of late-type stars observed in old clusters \citep{Meibom2005}. However, resonance locking can only explain circularization periods up to 10 days at the end of the main-sequence and requires unrealistic slow stellar rotation to explain the observed circularization periods of old Halo stars up to 16 days \citep[see also the review of resonance locking by][]{Savonije2008}. \citet{Ogilvie2007} found that inertial waves in the convective envelope excited by the Coriolis force can significantly enhance tidal dissipation in solar-like stars (although only up to $P\sim 10\,$d). Tidal dissipation in stellar interiors is still an unresolved subject due to the poorly understood and difficult-to-model physical processes that govern convection, differential rotation, magnetism, the coupling of different oscillation modes and their interactions with tidal flows \citep[see eg.][]{Goldman1991, Penev2007, Goldman2008, Ogilvie2012, Vidal2020, Barker2020, ZanazziWu2021}. Currently there is no consensus on a unified tidal theory that explains the efficient circularization of binaries with large orbital distances within reasonable assumptions \citep[see e.g. reviews by][]{Savonije2008, Mazeh2008, Zahn2008}. 

\subsection{Companions} \label{ssec:companions}
Finally the orbital evolution of a binary may have been affected by the presence of additional stellar companions. \citet{Tokovinin2006} found that the fraction of binary systems with additional companions is a sharp function of orbital period. For binaries with $P < 3$ days, $96\%$ of systems have an outer companion while only $34\%$ of binaries with $P>12$ days have additional companions. This strongly indicates that multi-body dynamics have affected the orbits of the closest binaries. An outer third body will typically increase the eccentricity of the inner pair {\citep[e.g. via the Kozai-Lidov mechanism, see][]{2007ApJ...669.1298F,2013ApJ...763L...2D, 2016ARA&A..54..441N}.} The increase in eccentricity will bring the inner pair close at periastron. Tidal forces acting at periastron can then circularise the orbit, resulting in a circularized, close binary. However, such an interaction still requires enhanced tidal forces to circularise binaries at large orbital separations.

\subsection{Tides in Exoplanetary Systems}
Tides act not only in double stars but also in exoplanetary systems. \citet{2017A&A...602A.107B} recently performed a homogeneous orbital analysis of 231 transiting giant planets and found by analysing eccentricities that tides have shaped the orbits of virtually all giant planets with orbital separations $a/R \lesssim 100$. Tides act to not only circularise orbits but also to modify stellar spin rates and realign stellar spin axes. \citet{2018AJ....155..165P} found evidence that cool stars hosting hot Jupiters have been tidally spun up. The obliquity (spin-orbit alignment angle) has been measured for a large sample of transiting hot Jupiters. Measurements of the projected obliquity of systems harboring hot Jupiters have shown that cool host stars display good spin-orbit alignment while hot Jupiters orbiting hot host stars often have large obliquities. The change in the observed spin-orbit distribution appears near the Kraft break at $6250\,$K at the disappearance of the convective envelope \citep{Winn2010_obl}. This has been interpreted as evidence that tides have aligned the orbits of these planets \citep{Winn2010_obl, Albrecht2012}. More recently it appears as if a simple two-population model cannot explain the observed obliquity distribution of hot Jupiters \citep{2019ESS.....420201A, 2019ESS.....420202J}. While the majority of hot Jupiters below $6250\,$K are aligned, a mixture of aligned and misaligned systems are observed in the range $\sim\! 6200-8000\,$K. At hotter temperatures nearly all hot Jupiters are misaligned. This indicates that tides work more efficiently in some stars in this temperature range than predicted, in qualitative agreement with our observed trends for binary stars of similar temperatures.

\section{Conclusions} \label{sec:conclusions}
We presented a catalog of eclipsing binaries in the southern elliptical hemisphere observed by TESS. We modelled light curves and spectral energy distributions to obtain orbital and stellar parameters for this sample and validated the accuracy against known samples. As a first application of our catalog, we combined TESS binaries with \textit{Kepler} and double star systems from the catalog compiled by \cite{Torres2010} to study tidal circularization of close binaries. Our main observational findings are: 
\begin{enumerate}
    \renewcommand{\labelenumi}{\roman{enumi})}
    \item The onset (and magnitude) of eccentricity depends on orbital distance {\it and} stellar temperature.
    \item The fraction of binaries with eccentric orbits increases with orbital distance from zero to one for all temperatures{, \textit{i.e.} all binaries at large orbital distances are found on eccentric orbits.}
\end{enumerate}

\noindent How do these findings compare to theory?

\begin{enumerate}    
    \renewcommand{\labelenumi}{\alph{enumi})}
    \item It is assumed that binaries form with a primordially broad eccentricity distribution and that circular orbits are most often a result of tidal circularization. This is consistent with the observation that the eccentricity fraction increases with $a/R_1$ from zero to one for all binaries.
    \item The eccentricity distribution of convective binaries is in excellent agreement with the predictions from the equilibrium tide. We do however observe some eccentric systems with small orbital distances, likely due to orbital perturbations from third companions in these systems. We similarly observe a small fraction of systems with circular orbits at larger orbital separations than predicted. These systems may have formed with primordially circular orbits or experienced enhanced tidal circularization.
    \item While the observed onset of eccentricity of hot binaries {($T_{\rm eff}\gtrsim6250\,$K)} at $a/R_1 \sim 3$ is in good agreement with the prediction from the dynamical tide, we observe a large fraction circularised binaries at distances of $a/R_1 \sim 3-10$ with {temperatures} in the range $6250-10,000\,$K, hinting at a population of hot binaries which have experienced significantly enhanced tidal dissipation. The eccentricity distribution of binaries hotter than $10,000\,$K is in excellent agreement with theory. 
\end{enumerate}

The effort presented in this paper is only a first look at eclipsing binaries with TESS. A homogeneous search and analysis of the full TESS data search is guaranteed to reveal many more eclipsing binaries that will offer further observational constraints on tidal theory and many other subjects in stellar physics. The eclipsing binaries observed by TESS are ideal for radial velocity follow-up, enabling detailed modelling of individual systems of interest.\\

We thank Emil Knudstrup for computing the power spectra and TLS periodograms that formed the starting point of our eclipsing binary search. We thank Diana Windemuth for supplying stellar effective temperatures of \textit{Kepler} eclipsing binaries. {We thank the anonymous referee for helpful comments which improved the clarity of our manuscript.} We acknowledge support from the Danish Council for Independent Research, through a DFF Sapere Aude Starting Grant no. 4181-00487B. Funding for the Stellar Astrophysics Centre is provided by The Danish National Research Foundation (Grant agreement no.: DNRF106). This paper includes data collected by the TESS mission. Funding for the TESS mission is provided by the NASA Explorer Program. This work has made use of data from the European Space Agency (ESA) mission
{\it Gaia} (\url{https://www.cosmos.esa.int/gaia}), processed by the {\it Gaia}
Data Processing and Analysis Consortium (DPAC,
\url{https://www.cosmos.esa.int/web/gaia/dpac/consortium}). Funding for the DPAC
has been provided by national institutions, in particular the institutions
participating in the {\it Gaia} Multilateral Agreement. This research has made use of the SIMBAD database, operated at CDS, Strasbourg, France \citep{2000A&AS..143....9W}. \software{NumPy \citep{van2011numpy}, SciPy \citep{scipy}, Astropy \citep{2018AJ....156..123T}, batman \citep{Kreidberg2015}, Lightkurve \citep{2018ascl.soft12013L}, Astroquery \citep{2019AJ....157...98G}, pysynphot \citep{2013ascl.soft03023S}, emcee \citep{ForemanMackey2013}, corner.py \citep{corner}, matplotlib \citep{Hunter:2007}.}

\bibliographystyle{aasjournal}
\bibliography{TESSEB.bib}

\end{document}